\documentclass[final,3p,times,twocolumn]{elsarticle}
\usepackage{epsfig}
\usepackage{subfigure}
\usepackage{graphicx}
\usepackage{bm}
\usepackage{color}

\usepackage{multirow}
\usepackage{array}
\usepackage{setspace}

\usepackage{amssymb}
\usepackage{amsthm,amsmath}

\journal{Physica D}

\begin{document}

\begin{frontmatter}

\title {Nonlinear Ring Waves in a Two-Layer Fluid }

\author[label1]{K.R. Khusnutdinova*}
\address[label1]{Department of Mathematical Sciences, Loughborough University, Loughborough LE11 3TU, UK\\
*Corresponding author. Electronic mail: K.Khusnutdinova@lboro.ac.uk.}

\author[label1]{X. Zhang}


\begin{abstract}
Surface and interfacial weakly-nonlinear ring waves in a two-layer fluid are modelled numerically, within the framework of the recently derived  2+1-dimensional cKdV-type equation. In a case study, we consider concentric waves from a localised initial condition and waves in a 2D version of the dam-break problem, as well as discussing the effect of a piecewise-constant shear flow. The modelling shows, in particular, the formation of 2D dispersive shock waves (DSWs) and oscillatory wave trains. The surface and interfacial DSWs generated in our numerical experiments look distinctively different.

\end{abstract}





\end{frontmatter}

\bigskip

\section{Introduction}
The cylindrical (or concentric) Korteweg - de Vries (cKdV) equation 
\begin{equation}
2 A_R+3 A A_\xi +\frac13 A_{\xi\xi\xi}+ \frac{A}{R}=0
\label{classicalckdv}
\end{equation}
was derived and studied  in various physical contexts (see, for example, \cite{Maxon74} - \cite{Johnson_book} and references therein).
In particular, it was derived to describe surface waves in a uniform fluid from Boussinesq equations \citep{Miles78}  and from the set of Euler equations  \citep{Johnson80}. The cKdV-type equations were also derived for internal waves in a stratified fluid without shear flow \citep{Lipovskii85}, and surface waves in a uniform fluid with a shear flow \citep{Johnson90}. 
Recently, we generalised these studies by considering the propagation of internal and surface ring waves in a stratified fluid over a shear flow \citep{kx14}. The derived 2+1-dimensional cKdV-type equation has the form
\begin{equation}
\mu_1 A_R + \mu_2 A A_{\xi} + \mu_3 A_{\xi \xi \xi} + \mu_4 \frac{A}{R} + \mu_5 \frac{A_\theta}{R} = 0. \label{ggcKdV}
\end{equation}
Importantly, the coefficient $\mu_5$ is equal to zero both when the fluid is uniform and when there is no shear flow. In these cases the equation (\ref{ggcKdV}) reduces to a 1+1-dimensional model \citep{kx14}. The general theory was then applied to the case of a two-layer fluid with a piecewise-constant shear flow, with an emphasis on the analytical description of the wavefronts of surface and interfacial ring waves. We also note that the original cKdV equation ({\ref{classicalckdv}}) is integrable \citep{Druma76, Calogero78}.

In this paper, we use the derived model equations to study surface and interfacial ring waves in a two-layer fluid numerically.  The paper is organised as follows. In Section II, we briefly overview the derivation of the cKdV-type equation from the Euler equations with application to the two-layer fluid given in \cite{kx14}. Particular examples of concentric waves generated from a localised initial condition and a 2D version of the  dam-break problem are modelled  in Section III. The effect of a piecewise-constant  shear flow on the ring waves generated from a localised initial condition is modelled in section IV. Some conclusions are drawn in Section V.  The coefficients of the relevant cKdV-type equations are listed in Appendix A. The derivation of the 2D linear wave equations for the surface and interfacial waves is discussed  in Appendix B. The finite-difference scheme used in our study is described in Appendix C.


\section{Derivation of a cKdV-type equation}
\subsection{Problem formulation and amplitude equation}
We study a ring wave propagating in an inviscid incompressible fluid, described by the set of Euler equations:
\begin{eqnarray}
&& \rho (u_t + u u_x + v u_y + w u_z) + p_x = 0, \label{1} \\
&& \rho (v_t + u v_x + v v_y + w v_z) + p_y = 0, \label{2} \\
&& \rho (w_t + u w_x + v w_y + w w_z) + p_z + \rho g = 0, \label{3} \\
&& \rho_t + u \rho_x + v \rho_y + w \rho_z = 0, \label{4} \\
&& u_x + v_y + w_z = 0, \label{5} 
\end{eqnarray}
with the free surface and rigid bottom boundary conditions appropriate for the oceanic applications:
\begin{eqnarray}
&&w = h_t + u h_x + v h_y \quad \mbox{at} \quad z = h(x,y,t), \label{6} \\
&&p = p_a \quad \mbox{at} \quad z = h(x,y,t), \label{7} \\
&&w = 0 \quad \mbox{at} \quad z = 0. \label{8}
\end{eqnarray}
Here, $u,v,w$ are the velocity components in $x,y,z$ directions respectively, $p$ is the pressure, $\rho$ is the density, $g$ is the gravitational acceleration, $z = h(x,y,t)$ is the free surface height (with $z = 0$ at the bottom), and $p_a$ is the constant atmospheric pressure at the surface. We assume that in the basic state $u_0 = u_0(z), ~ v_0 = w_0 = 0, ~p_{0z} = - \rho_0 g, ~h = h_0$. Here $u_0(z)$ is a horizontal  shear flow in the $x$-direction, and  $\rho_0 = \rho_0(z)$ is a stable background density stratification.
	
We use the vertical particle displacement $\zeta$ as an additional dependent variable, which is defined by the equation
\begin{equation}
\zeta_t + u \zeta_x + v \zeta_y + w \zeta_z = w, \label{9}
\end{equation}
and satisfies the surface boundary condition 
\begin{equation}
\zeta = h - h_0 \quad  \mbox{at} \quad z = h(x,y,t), \label{10}
\end{equation}
where $h_0$ is the unperturbed depth of the fluid.

We use the following non-dimensional variables:
\begin{eqnarray*}
&&x \to \lambda x, \quad y \to \lambda y, \quad z \to h_0 z, \quad t \to \frac{\lambda}{c^*}t, \\
&& u \to c^* u, \quad v \to c^* v, \quad w \to \frac{h_0 c^*}{\lambda} w, \\
&&(\rho_0, \rho) \to \rho^*( \rho_0, \rho),  \quad h \to h_0 + a \eta, \\
&& p \to p_a + \int_{z}^{h_0} \rho^* \rho_0(s) g ~\mathrm{d} s + \rho^* g h_0 p,
\end{eqnarray*}
where $\lambda$ is the wave length, $a$ is the wave amplitude, $c^*= \sqrt{g h_0}$ is the  long-wave speed of surface waves, 
$\rho^*$ is the dimensional reference density of the fluid, while $\rho_0(z)$ is the non-dimensional function describing stratification in the basic state, and $\eta = \eta(x,y,t)$ is the non-dimensional free surface perturbation. 
Non-dimensionalisation leads to the appearance of two small parameters in the problem, the amplitude parameter $\varepsilon = a/h_0$ and the wavelength parameter $\delta = h_0/\lambda$.  For the sake of simplicity, in the subsequent derivation we impose the condition $\delta^2 = \varepsilon$.
Variables can be scaled further to replace $\delta^2$ with $\varepsilon$ in the equations  \citep{Johnson_book}. 


We introduce the cylindrical coordinate system moving at a constant speed $c$ (a natural choice is the flow speed at the bottom, as follows from the derivation) and consider deviations from the basic state (the same notations $u$ and $v$ have been used for the projections 
on the new coordinate axis), scaling the appropriate variables by the amplitude parameter $\varepsilon$,
 \begin{eqnarray*}
&x \to ct + r \cos \theta, ~~ y \to r \sin \theta, ~~ z \to z, ~~ t \to t, \\
&u \to u_0(z) + \varepsilon (u \cos \theta - v \sin \theta), \\
& v \to \varepsilon (u \sin \theta + v \cos \theta), \\
 &w \to \varepsilon w, ~~ p \to  \varepsilon p, ~~ \rho \to \rho_0 + \varepsilon \rho.
 \end{eqnarray*}
 
 Then, we look for a solution of the problem in the form of asymptotic multiple-scales expansions of the form
 $
 \zeta = \zeta_1 + \varepsilon \zeta_2 + \dots,
 $
 and similar expansions for other variables, where
 \begin{equation}
 \zeta_1 = A(\xi, R, \theta) \phi(z, \theta), \label{11}
 \end{equation}
 with the following set of fast and slow variables:
 \begin{eqnarray}
 \xi = r k(\theta) - s t, \quad R = \varepsilon r k(\theta), \quad \theta = \theta,
 \end{eqnarray}
 where  we define $s$ to be the wave speed in the absence of a  shear flow (with $k(\theta) = 1$), while when a shear flow is present the function $k(\theta)$ describes the distortion of the wavefront in a particular direction, and is to be determined.  
 The formal range of asymptotic validity of the model is defined by the conditions $\xi\sim R\sim O(1)$.  To leading order, the wavefront at any fixed moment of time $t$ is described by the equation
$rk(\theta)=\mbox{constant},$
 and we consider  outward propagating ring waves, requiring that 
 $k = k(\theta) > 0$.
 
 To leading order, assuming that perturbations of the basic state are caused only by the propagating wave, we obtain 
 \begin{eqnarray}
&& u_1 = - A \phi u_{0z} \cos \theta - \frac{k F}{k^2 + k^{'2}} A\phi_z, \label{O1_1} \\
&& v_1 = A \phi u_{0z} \sin \theta -  \frac{k' F}{k^2 + k^{'2}} A\phi_z,  \label{O1_2}\\
&& w_1 =  A_{\xi} F \phi,  \label{O1_3} \\
&& p_1 = \frac{\rho_0}{k^2 + k^{'2}} A F^2 \phi_z,  \label{O1_4}  \\
&& \rho_1 = - \rho_{0z} A \phi,  \label{O1_5}  \\
&&\eta_1 = A \phi \quad \mbox{at} \quad z = 1,  \label{O1_6} 
\end{eqnarray}
where the function $\phi = \phi(z, \theta)$ satisfies the following modal equations:
\begin{eqnarray}
&\left (\frac{\rho_0 F^2}{k^2 + k^{'2}} \phi_z\right )_z -  \rho_{0z} \phi = 0, \label{m1} \\
&\frac{F^2}{k^2 + k^{'2}} \phi_z -  \phi = 0 \quad \mbox{at} \quad z=1, \label{m2} \\
&\phi = 0 \quad \mbox{at} \quad z=0, \label{m3}\\
 \mbox{and }~&  F = -s + (u_0 - c) (k \cos \theta - k' \sin \theta),\nonumber
\end{eqnarray}
 where we now have fixed the speed of the moving coordinate frame $c$ to be equal to the speed of the shear flow at the bottom, $c =u_0(0)$. Then, $F = -s \ne 0$ at $z = 0$, and the condition $F \phi = 0$ at $z=0$ implies (\ref{m3}), simplifying the mathematical formulation. 
 The values of the wave speed $s$ in the absence of the shear flow, and the pair of functions $\phi(z, \theta)$ and $k(\theta)$,  for a given shear flow,  constitute solution of the modal equations (\ref{m1}) - (\ref{m3}).  
 
 Substituting the leading order solution (\ref{O1_1})-(\ref{O1_6}) into the equations at order $O(\varepsilon)$, we obtain the following non-homogeneous equation for the function  $\zeta_2$;
  \begin{equation}
 \left ( \frac{\rho_0 F^2}{k^2 + k'^2} \zeta_{2\xi z} \right )_z - \rho_{0z} \zeta_{2\xi} = M_2, \label{NE}
 \end{equation}
   with the boundary conditions 
   \begin{equation}
 \zeta_{2 \xi} = 0 \quad \mbox{at} \quad z=0, \label{NBC1}
\end{equation}
    \begin{equation}
 \rho_0 \left [ \frac{F^2}{k^2 + k'^2} \zeta_{2\xi z} - \zeta_{2\xi} \right ] = N_2 \quad \mbox{at} \quad z=1, \label{NBC2}
 \end{equation}
where $M_2$ and $N_2$ are explicitly given in terms of the solutions of the leading order problem (see \cite{kx14}).
The compatibility condition
$
\int_0^1 M_2 \phi ~ \mathrm{d}z - [N_2 \phi]_{z=1} = 0
$
yields the 2+1-dimensional evolution equation for the slowly varying amplitude of the ring wave in the form
\begin{equation}
\mu_1 A_R + \mu_2 A A_{\xi} + \mu_3 A_{\xi \xi \xi} + \mu_4 \frac{A}{R} + \mu_5 \frac{A_\theta}{R} = 0.\label{cKdV}
\end{equation}
The coefficients are given in terms of the solutions of the modal equations (\ref{m1}) - (\ref{m3}) by the formulae:
\begin{align}
\mu_1 = 2 s \int_0^1 \rho_0 F \phi_z^2 ~ \mathrm{d}z,
\quad\quad\qquad\qquad\qquad\qquad\qquad\qquad 
\label{c1}\\
\mu_2 = - 3 \int_0^1 \rho_0 F^2 \phi_z^3 ~ \mathrm{d}z, ~
\quad\qquad\qquad\qquad\qquad\qquad\qquad
\label{c2}\\
\mu_3 = - (k^2 + k'^2) \int_0^1  \rho_0 F^2 \phi^2 ~ \mathrm{d}z, ~
\qquad\qquad\qquad\qquad\qquad  
\label{c3}\\
\mu_4 = - \int_0^1 \bigg(\frac{\rho_0 \phi_z^2 k (k+k'')}{(k^2+k'^2)^2} \big ( (k^2-3k'^2) F^2 ~
\qquad \qquad\qquad \quad 
\nonumber \\
 -4 k' (k^2 + k'^2) W_0 F \sin \theta     - W_0^2(k^2 + k'^2)^2  \sin^2 \theta \big ) 
 \qquad 
 \nonumber\\
 +  \frac{2 \rho_0 k}{k^2 + k'^2} F \phi_z \phi_{z\theta} (k' F + (k^2 + k'^2) W_0 \sin \theta ) \bigg) ~ \mathrm{d}z,
 \quad \quad 
 \label{c4}\\
  \mu_5 = - \frac{2k}{k^2 + k'^2} \int_0^1 \rho_0 F \phi_z^2 [k' F + W_0 (k^2+k'^2) \sin \theta ]~\mathrm{d}z, 
  \qquad 
   \label{c5}
\end{align}
where $W_0 = u_0-c$.

 \subsection{Two-layer fluid}
 

We consider the case when both the density of the fluid and the shear flow are piecewise-constant functions ($0 \le z \le 1$):	
\begin{eqnarray*}
&&\rho_0 = \rho_2 H(z) + (\rho_1 - \rho_2) H(z-d), \\
&&u_0 = U_2 H(z) + (U_1 - U_2) H(z-d).
\end{eqnarray*}
Here, $d$ is the thickness of the lower layer and $H(z)$ is the Heaviside function.  This background flow is subject to Kelvin-Helmholtz instability, which is excluded in the consideration of a long wave over a sufficiently weak shear flow (see \cite{kx14} for a relevant discussion and the references). 
 
 Solution of the modal equations  (\ref{m1}) - (\ref{m3}) in the  upper and the lower layers is given, respectively, by
\begin{eqnarray}
&&\phi_1 = \left ( \frac{F_1^2}{k^2 + k'^2} + z - 1 \right ) {\Lambda}, \nonumber\\
&&\phi_2 =  \left ( \frac{F_1^2}{k^2 + k'^2} + d - 1 \right ) \frac{\Lambda z}{d},\label{phi}
\end{eqnarray}
where $\Lambda$ is a parameter, and the function $\phi$ is continuous, while the jump condition 
$$
\frac{\left [ \rho_0 F^2 \phi_z \right ]}{k^2 + k'^2} = [\rho_0] \phi \quad \mbox{at} \quad z=d
$$
provides an equation for the function $k(\theta)$:
\begin{align}
&&(\rho_2 - \rho_1) d (1-d) (k^2 + k'^2)^2 + \rho_2 F_1^2 F_2^2\nonumber\\
&&- \rho_2 [d F_1^2 + (1-d) F_2^2] (k^2 + k'^2) = 0,
\label{k}
\end{align}
with $F_1 = -s + (U_1 - U_2) (k \cos \theta - k' \sin \theta), F_2 = -s$.
This  nonlinear first-order ordinary differential equation  is further generalisation of the Burns and generalised Burns conditions \citep{ Burns53, Johnson90}.  

First, we assume that there is no shear flow and find the wave speed $s$ by letting $U_1=U_2=0$, while $k=1$. The dispersion relation takes the  standard form
$$\rho_2 s^4 -\rho_2 s^2 + (\rho_2-\rho_1)d (1-d)=0.$$
So the wave speed in the absence of the shear flow is given by
${\displaystyle s^2=
\frac{1}{2} \left (1 \pm \sqrt{(2d-1)^2+4\rho_1/\rho_2 d(1-d)} \right )}$, 
where the upper sign should be chosen for the faster surface mode, and the lower sign for the slower internal mode.
 
 The general solution of the equation (\ref{k}) can be found in the form similar to the general solution of the generalised Burns condition \citep{Johnson90}, allowing us then to find the necessary singular solution relevant to the ring waves in a stratified fluid in parametric form \cite{kx14}:
\begin{equation}
k(a)=-\frac{aQ_a-2Q}{\sqrt{(Q_a-2a)^2+4b^2}},\label{solutionk}
\end{equation}
where if $\theta\in(0,\pi)$,  then
$b=\sqrt{Q-a^2},$
$$\theta=\left\{
\begin{array}{ll}
\arctan (-\frac{2\sqrt{Q-a^2}}{Q_a-2a})&\quad \mbox{if} \quad Q_a-2a<0,\\
\arctan (-\frac{2\sqrt{Q-a^2}}{Q_a-2a})+\pi&\quad \mbox{if} \quad Q_a-2a>0,
\end{array}\right.$$
while if  $\theta\in(\pi,2\pi)$, then
$b=-\sqrt{Q-a^2},$ 
$$\theta=\left\{
\begin{array}{ll}
\arctan (\frac{2\sqrt{Q-a^2}}{Q_a-2a})+\pi&\quad \mbox{if} \quad Q_a-2a>0,\\
\arctan (\frac{2\sqrt{Q-a^2}}{Q_a-2a})+2\pi&\quad \mbox{if} \quad Q_a-2a<0.
\end{array}\right.$$
Here,
\begin{eqnarray*}
Q=\frac{\rho_2 [d(-s+a(U_1-U_2))^2+(1-d)s^2] \pm \sqrt{\Delta}}{2(\rho_2-\rho_1)d(1-d)},\\
\Delta =  \rho_2^2 [d (-s+a(U_1-U_2))^2 - (1-d) s^2]^2 \\
+  4  \rho_1 \rho_2 d (1-d) s^2 [-s+a(U_1-U_2)]^2, 
\end{eqnarray*}
where the upper (lower) sign should be chosen for the interfacial (surface) wave.
The wave speed $s$, the modal function $\phi(z, \theta)$, and the function $k(\theta)$  are now determined. The coefficients of the equation (\ref{cKdV}) for the surface and interfacial mode are listed in Appendix A.

 
\section{Nonlinear propagation of concentric waves}

In this section we model the propagation of concentric waves in a two-layer fluid. Assuming that there is no shear flow, one can derive the 2D wave equations for the linear surface and interfacial waves (see Appendix B). These equations are used to describe the initial evolution of the weakly-nonlinear waves. Then, we solve the derived cKdV equations for both modes, using the solutions of the linear equations at $R = R_0$ as the necessary `initial' conditions. The numerical scheme is an extension of the scheme suggested in \citep{Feng98} (see Appendix C).
	
\subsection{A localised initial condition}
The 2D linear wave equation
$$A_{tt}-s^2(A_{xx}+A_{yy})=0,$$
has an exact solution describing waves from a localised condition at $t=0$ \cite{Dobrokhotov}:
\begin{equation}
A(x,y,t)=Q ~\mbox{Re} \bigg( \frac {1+ist/\nu} {((1+ist/\nu)^2+(x^2+y^2)/\nu^2)^{3/2}}\bigg).
\label{exactsol}
\end{equation}
Here,  $Q$ and $\nu$ are arbitrary constants.
\begin{figure}
\centering
	\includegraphics[width=0.6\columnwidth]{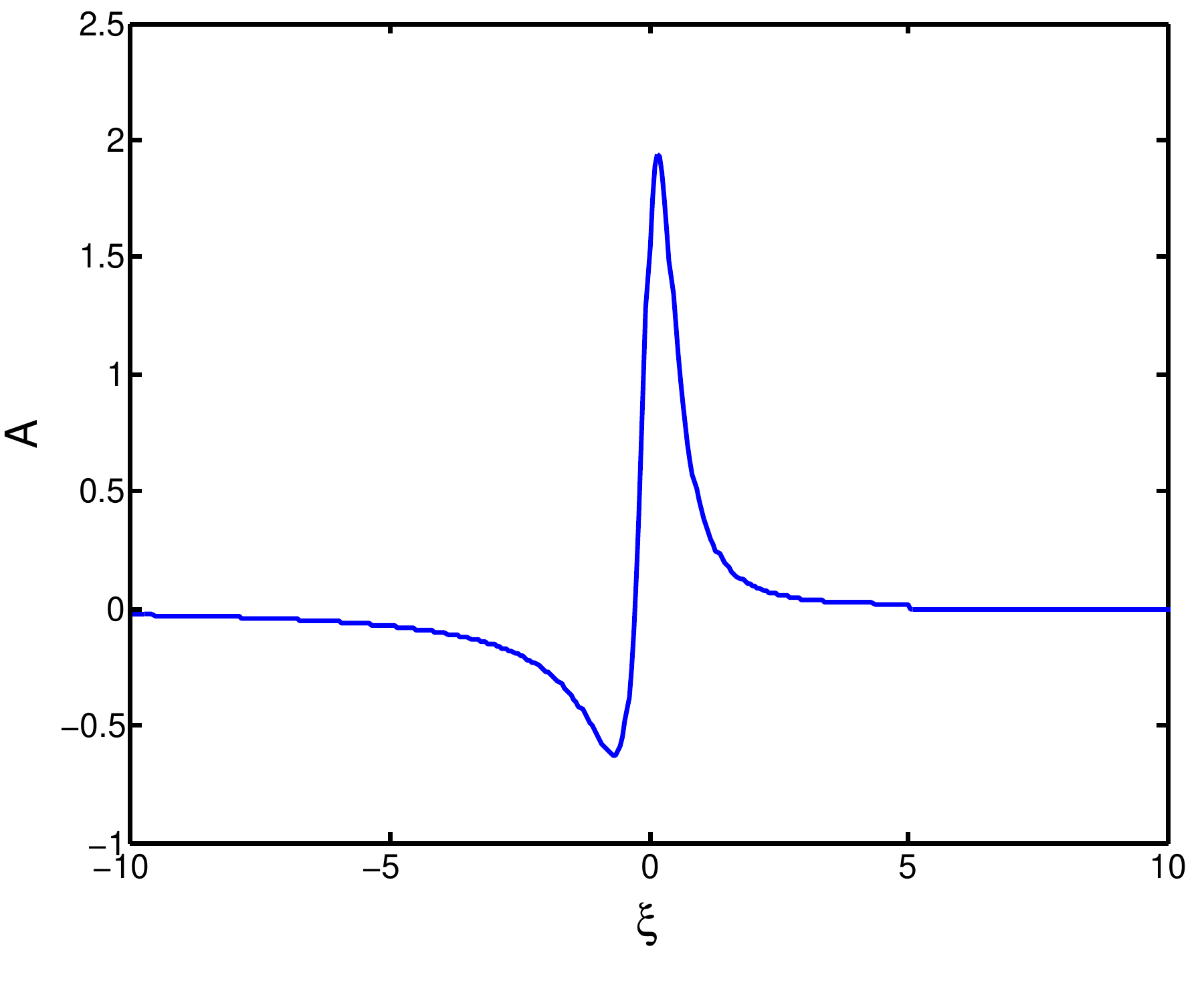} 
		\caption{Initial condition 
		at $R_0=0.1$ ($Q = 20, \nu = 0.5$).}
	\label{initialxi}
 \end{figure}

To solve the derived nonlinear cKdV equation numerically, we need to know 
the wave amplitude at $R_0=\varepsilon r_0$. In this example, we assume  that $\varepsilon=0.02$ and $\nu = 0.5$,  and choose the exact solution (\ref{exactsol}) written in $(\xi,R)$ coordinates as the initial condition at $R=R_0$:
$$A(R_0, \xi)= Q ~\mbox{Re} \left( \frac {1+2 i(50 R_0-\xi)} {((1+2i(50 R_0-\xi))^2+(100R_0)^2)^{3/2}}\right).$$
We use one and the same initial condition for both surface and interfacial waves. The initial condition at $R=R_0=0.1$ is shown in Figure \ref{initialxi}  for $Q=20$ and $\nu = 0.5$. We choose $\rho_{1}=1,~\rho_{2}=1.2$ and consider two values of the lower layer depth $d=0.5$ and $d=0.6$.

\subsubsection{Numerical results for surface waves}
 
\begin{figure}
\centering
\renewcommand{\thesubfigure}{(\arabic{subfigure})}
\subfigure[$Q=20$]{\includegraphics[width=.27\textwidth]{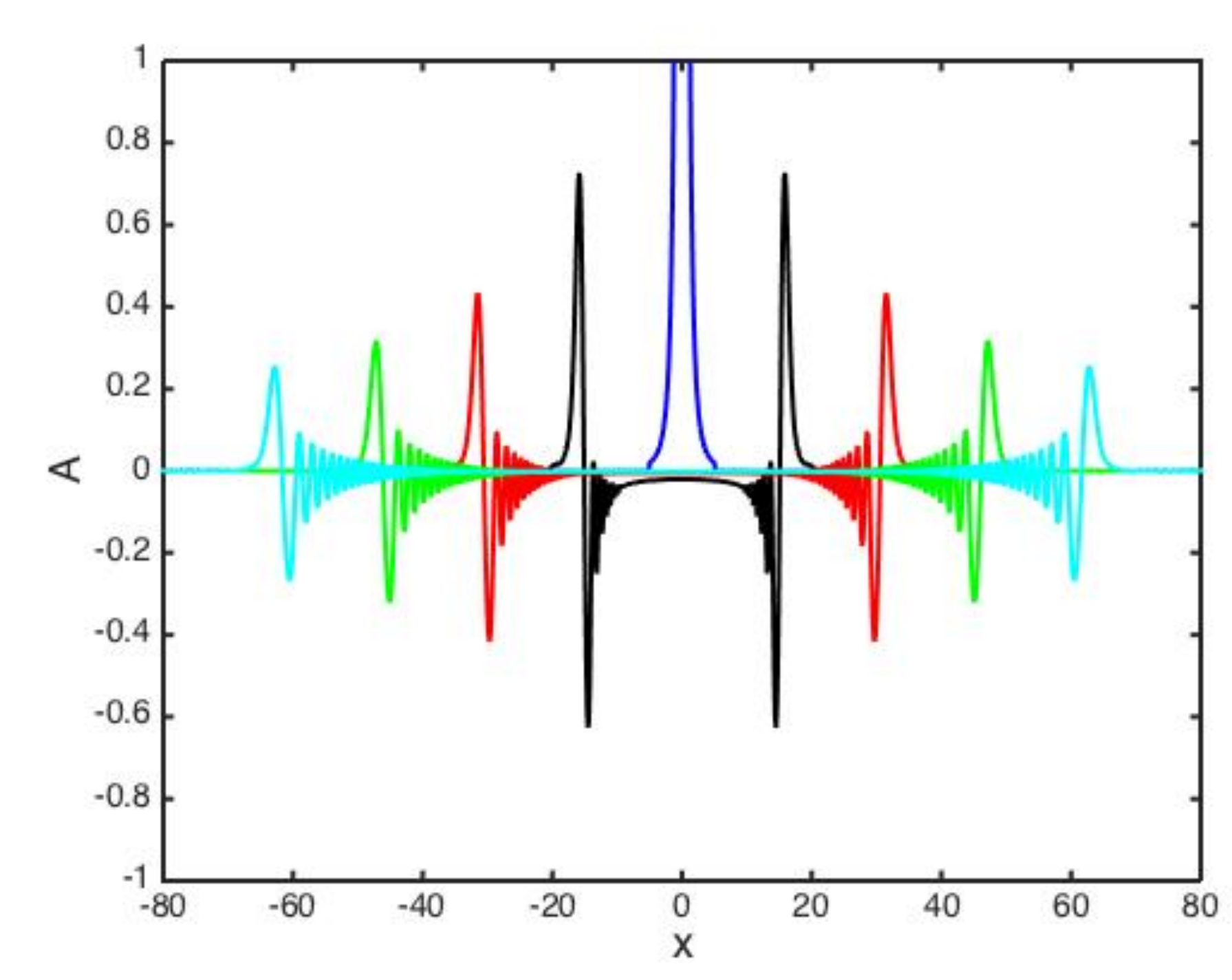}}
\subfigure[$Q=-20$]{\includegraphics[width=.27\textwidth]{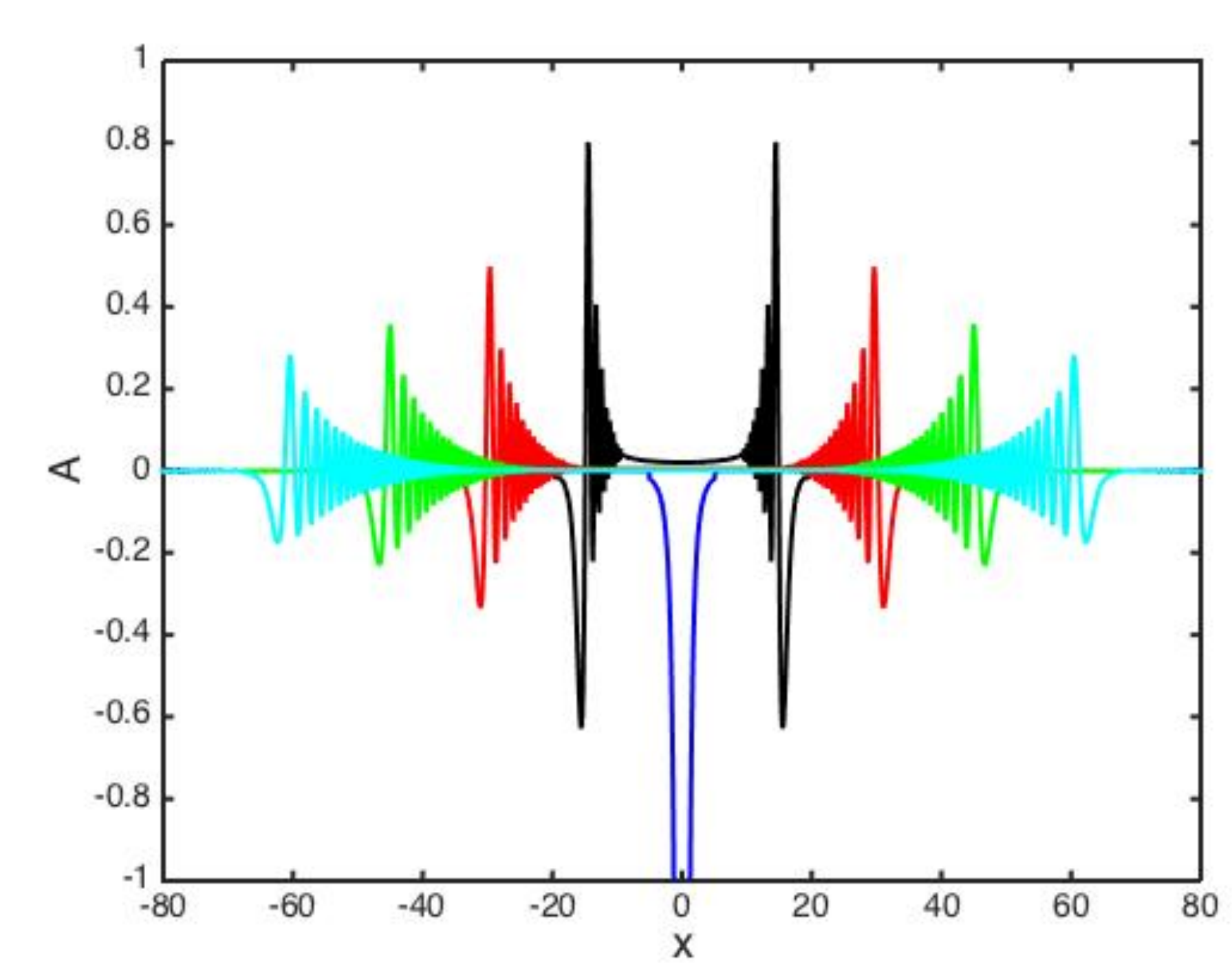}}
\caption{Surface waves in the directions $\theta = 0$ and $\theta = \pi$ for $d=0.6$ at
$t = 0, 16, 32, 48$ and $64$ (from a localised initial condition in the centre at $t=0$).
}
	\label{figure9b}
\setcounter{subfigure}{0}
\end{figure}

If  $d=0.6$, the wave speed of the surface waves is  $s_+\approx 0.9789$, and we solve the equation
\begin{align*}
2.1358A_R+3.2015AA_\xi+0.3236A_{\xi\xi\xi}+1.0679\frac{A}{R}=0.
\end{align*}
Numerical solutions are obtained for $Q=20$ and $Q=-20$. 
The cross-section along the directions $\theta=0$ and $\theta=\pi$ is shown in Figure \ref{figure9b}. 
The problem formulation and profiles of surface waves for $d=0.5$ are similar. 

\begin{figure}
\centering
\renewcommand{\thesubfigure}{(\arabic{subfigure})}
\subfigure[$Q=20$]{\includegraphics[width=.27\textwidth]{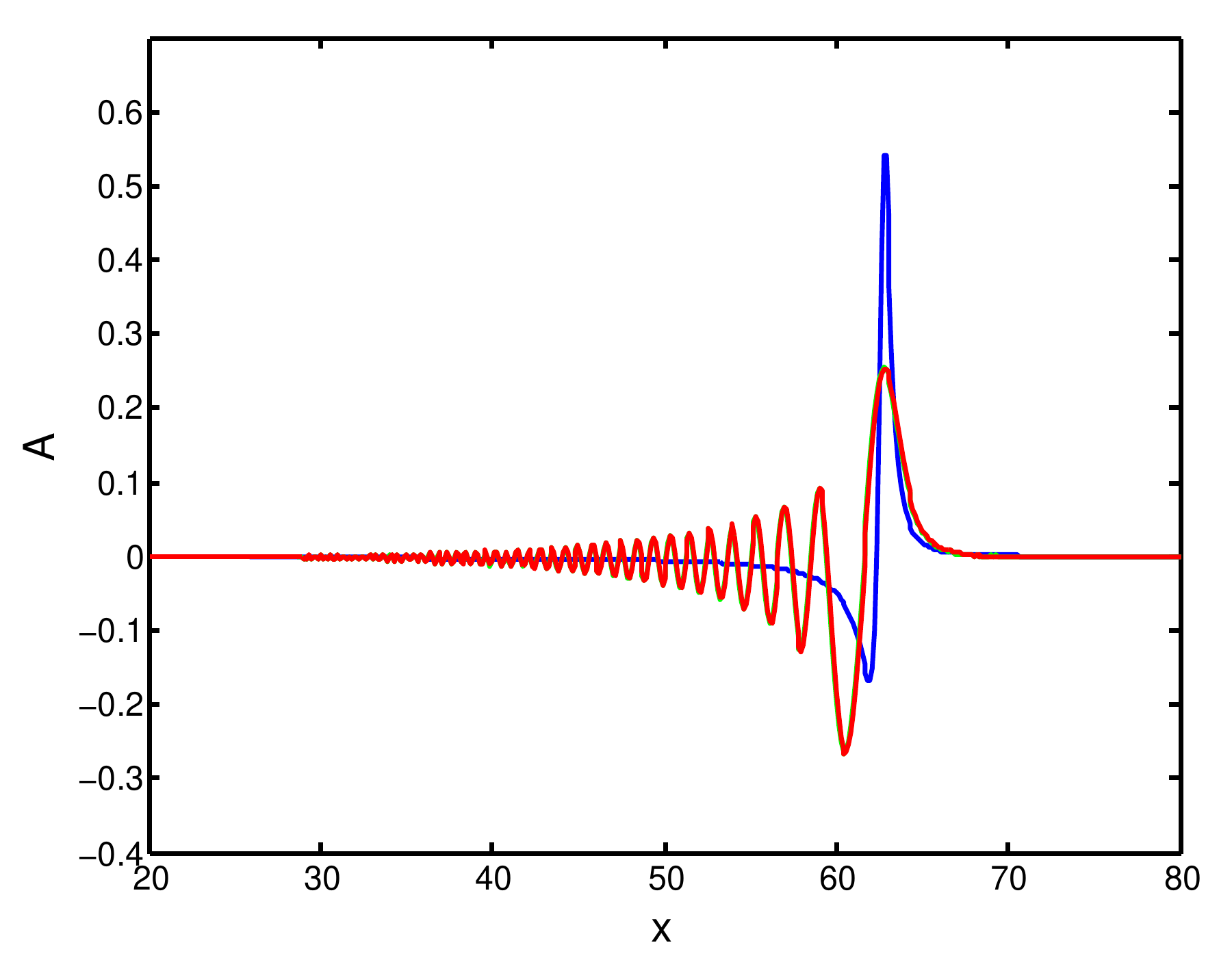}}
\subfigure[$Q=-20$]{\includegraphics[width=.27\textwidth]{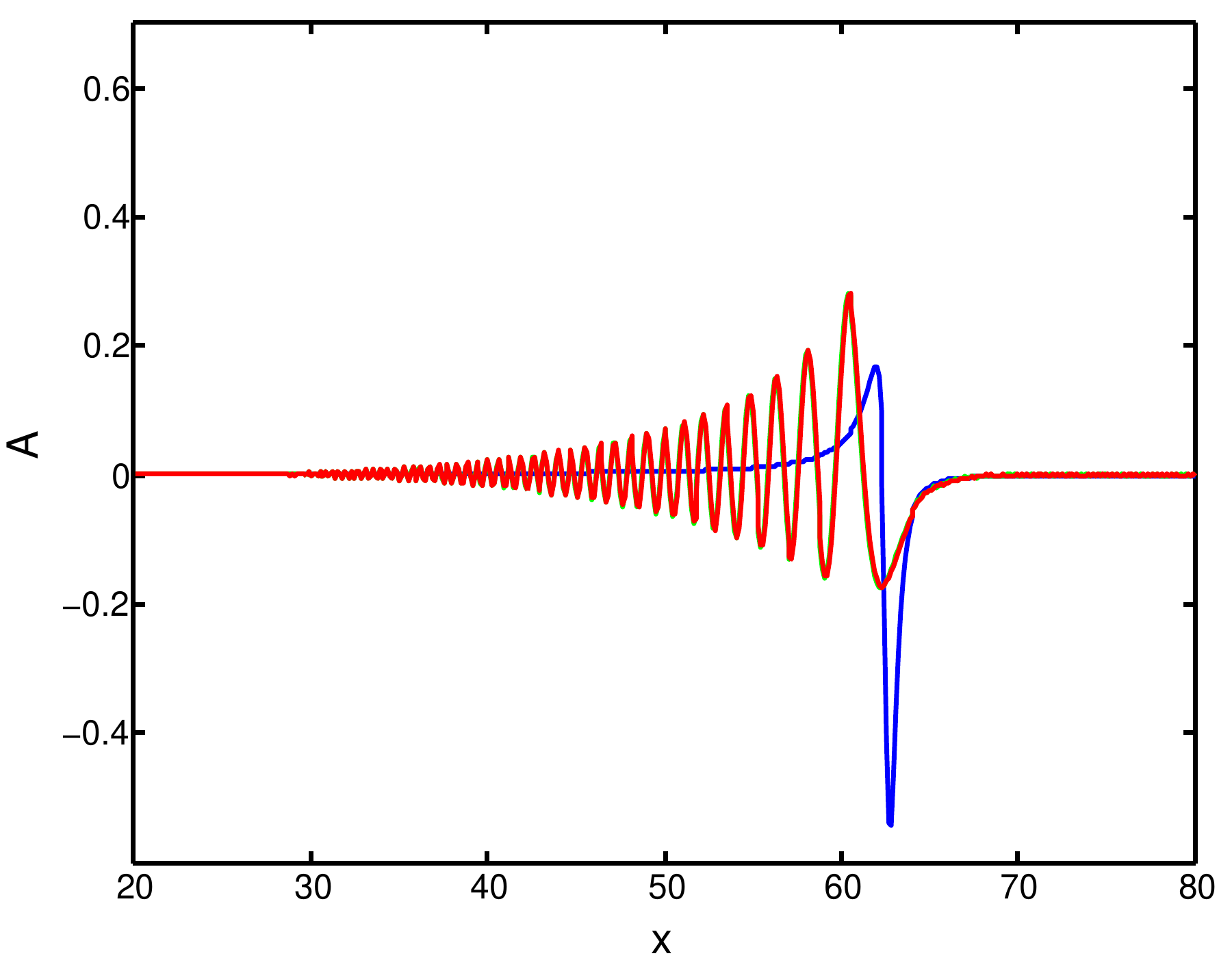}}
\caption{The linear (non-oscillatory)
and nonlinear (oscillatory)
surface waves for $d=0.6$ at $t=64$.}
	\label{figurelns}
\setcounter{subfigure}{0}
 \end{figure}

The exact linear and the numerical nonlinear solutions are compared in  Figure \ref{figurelns}  for $t=64$ and $d = 0.6$. Weak nonlinearity and dispersion, acting together, yield the generation of a well-developed  oscillatory dispersive wave train behind the lead wave of elevation ($Q > 0$) or depression ($Q < 0$), which is not captured by the 2D linear wave equation. The amplitude of the lead wave decreases with the increase of the distance from the centre much more rapidly than in the linear solution, which agrees with previous studies (e.g., \cite{Dorfman81, Stepanyants81}).

\subsubsection{Numerical results for interfacial waves}


\begin{figure}
\centering
\renewcommand{\thesubfigure}{(\arabic{subfigure})}
\subfigure[$Q=20$]{\includegraphics[width=.27\textwidth]{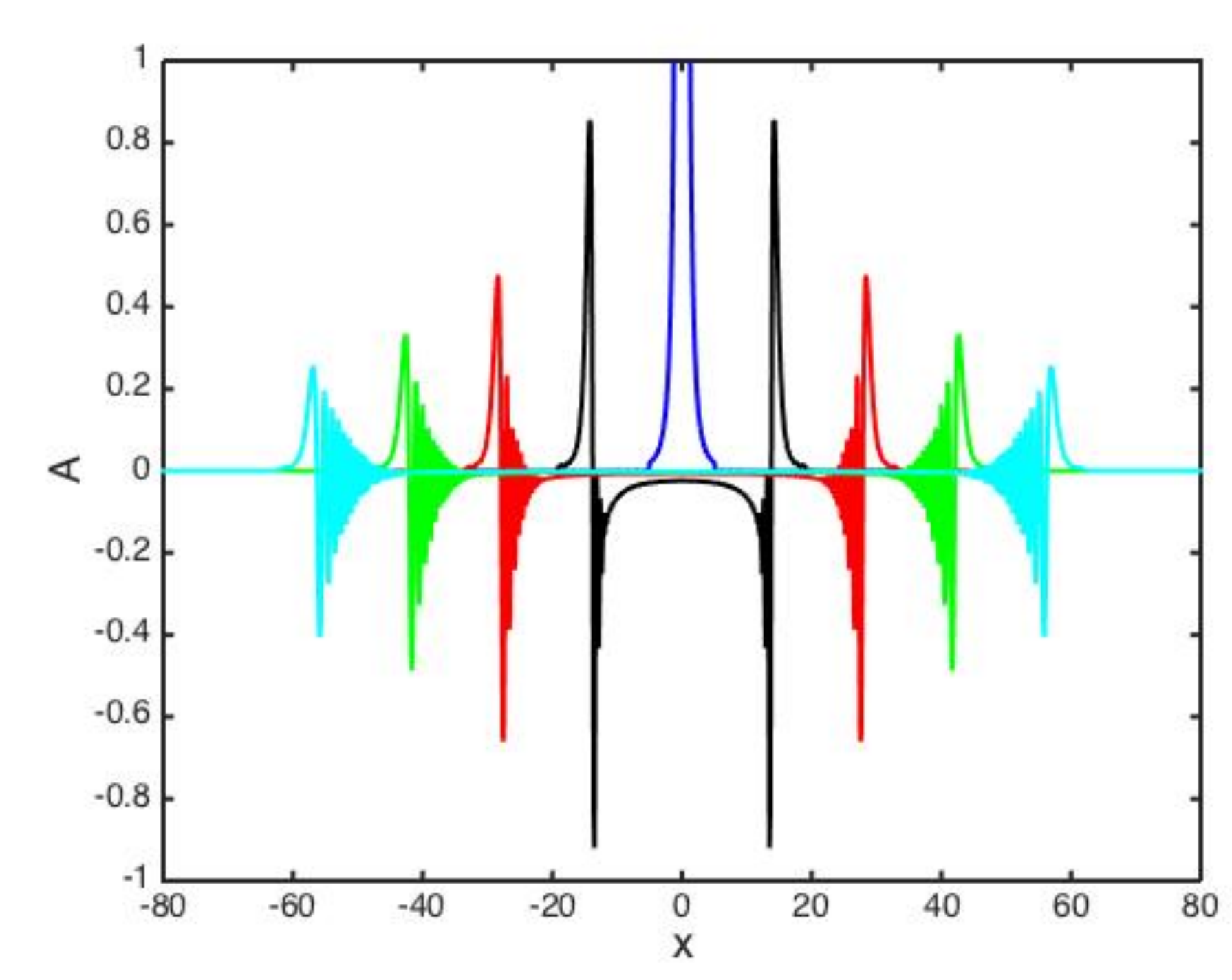}}
\subfigure[$Q=-20$]{\includegraphics[width=.27\textwidth]{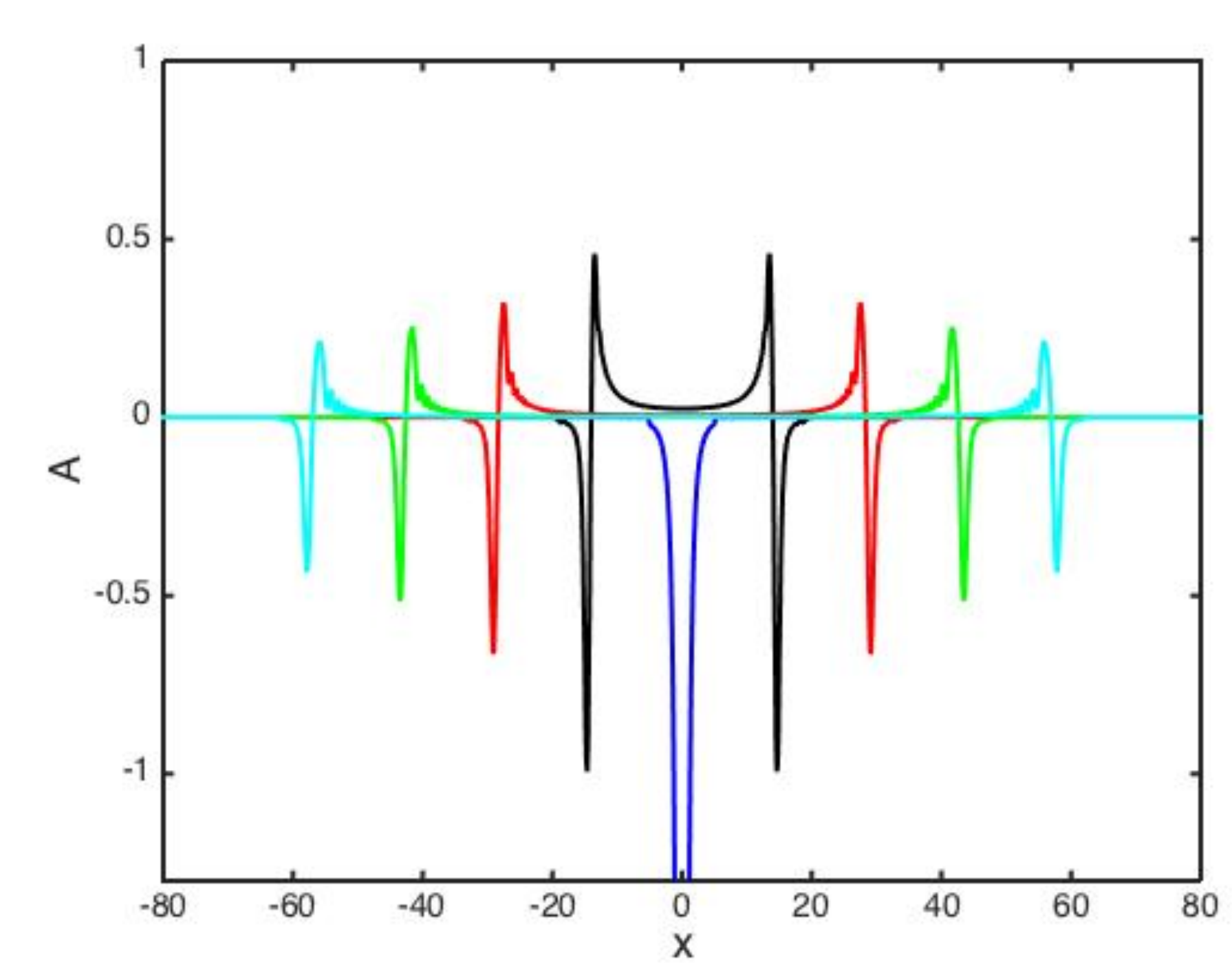}}
\caption{Interfacial waves in the directions $\theta = 0$ and $\theta = \pi$ for $d=0.6$ at 
$t=0, 70, 140, 210$ and $280$ (from a localised initial condition in the centre at $t=0$).
}
	\label{figure10b}
	\setcounter{subfigure}{0}
\end{figure} 

 \begin{figure}
 \centering
\renewcommand{\thesubfigure}{(\arabic{subfigure})}
\subfigure[$Q=20$]{\includegraphics[width=.27\textwidth]{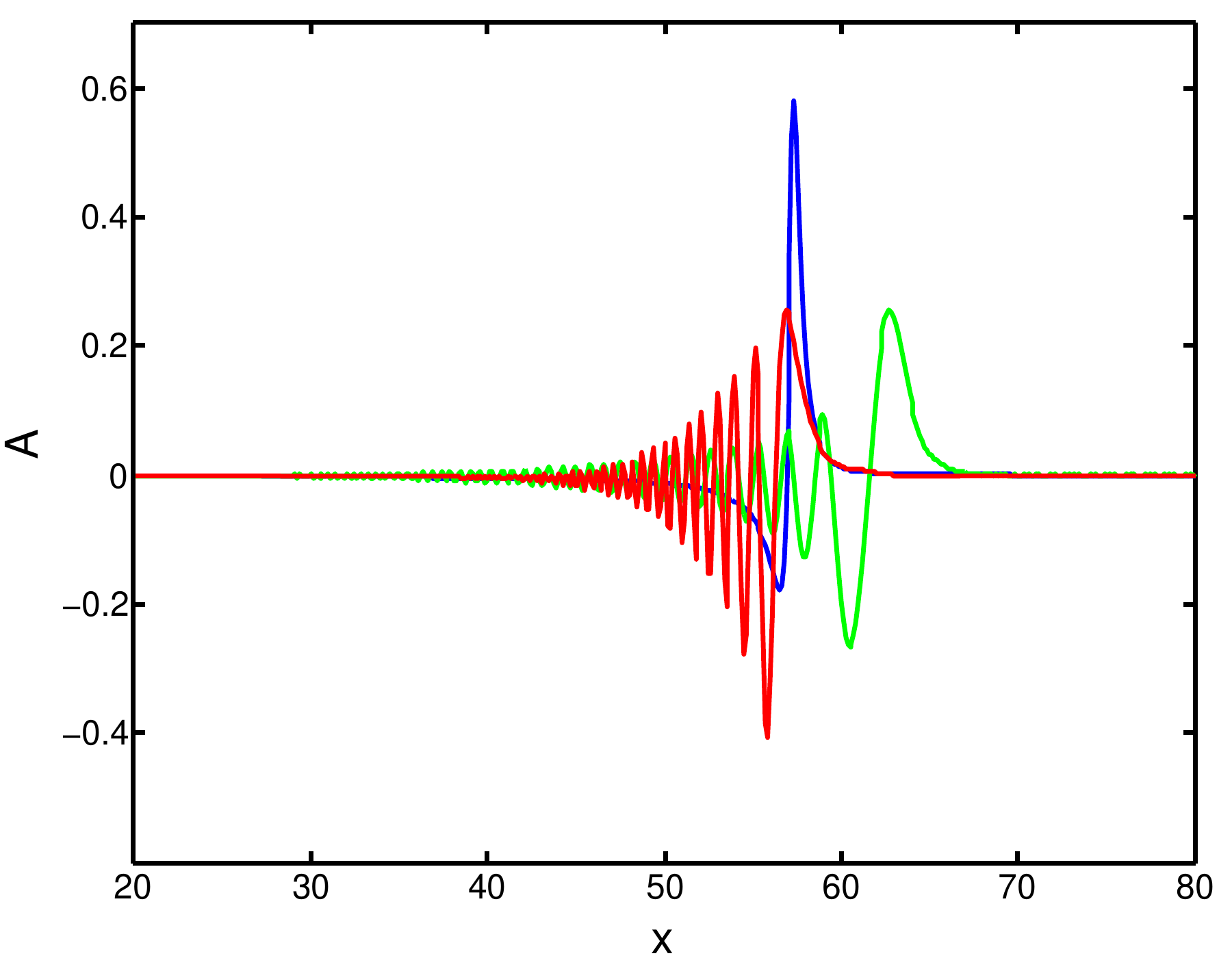}}
\subfigure[$Q=-20$]{\includegraphics[width=.27\textwidth]{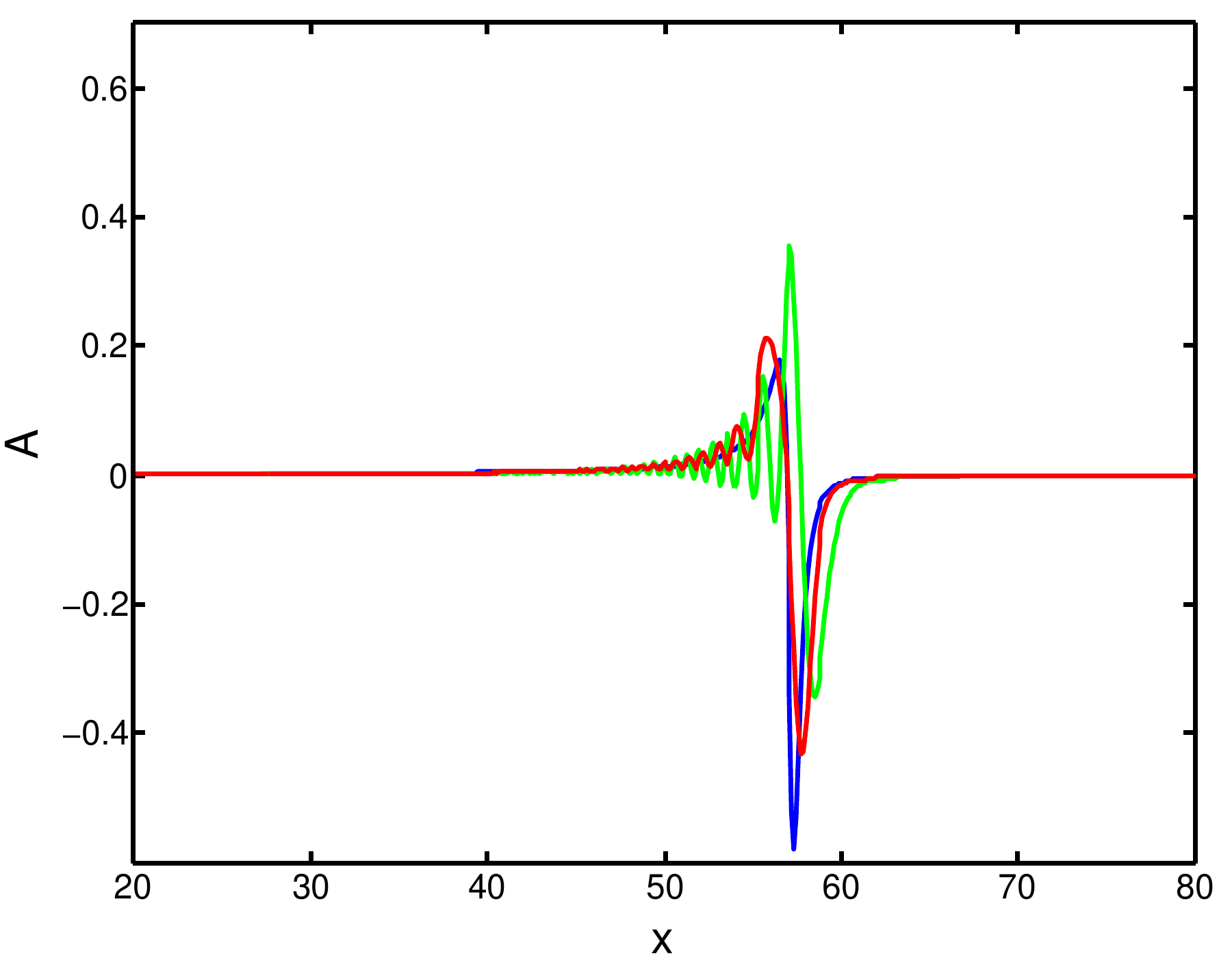}}
\caption{The linear (non-oscillatory)
and nonlinear (oscillatory) interfacial waves for $d=0.5$ (faster waves)
and $d=0.6$ (slower waves)
at $t=280$.}
\label{figurelni}
\setcounter{subfigure}{0}
 \end{figure}

If $d=0.5$, the wave speed of the interfacial waves is  $s_-\approx 0.2087$, and we solve  the equation
\begin{align*}
0.4182A_R-0.0599AA_\xi+0.0154A_{\xi\xi\xi}+0.2091\frac{A}{R}=0,
\end{align*}
while if $d=0.6$, then $s_-\approx 0.2043$, and the equation is given by
\begin{align*}
0.4272A_R-0.6719AA_\xi+0.0150A_{\xi\xi\xi}+0.2136\frac{A}{R}=0.
\end{align*}
Thus, the nonlinearity coefficient is much greater when $d = 0.6$, despite a small change in the thickness of the layers.

The numerical solutions are again shown for $Q=20$ and $Q=-20$. The cross-section of the wave profile  along the directions $\theta=0$ and $\theta=\pi$ is plotted 
in Figure \ref{figure10b}
for $d =0.6$.
The exact linear and the numerical nonlinear solutions for the interfacial wave are compared in Figure \ref{figurelni} for $t = 280$. When $Q>0$, the lead wave of elevation is more pronounced when the nonlinearity coefficient is small, while more energy goes into the formation of an oscillatory wave train in the second case. On the contrary, when $Q<0$, the lead wave of depression is more pronounced in the second case, and more energy goes into the formation of an oscillatory wave train in the first.  Thus, there are significant differences in the interfacial wave profiles, despite only a small change in the value of the parameter $d$.







\subsection{A 2D version of the dam-break problem}

\begin{figure}
\centering
	\includegraphics[width=0.55\columnwidth]{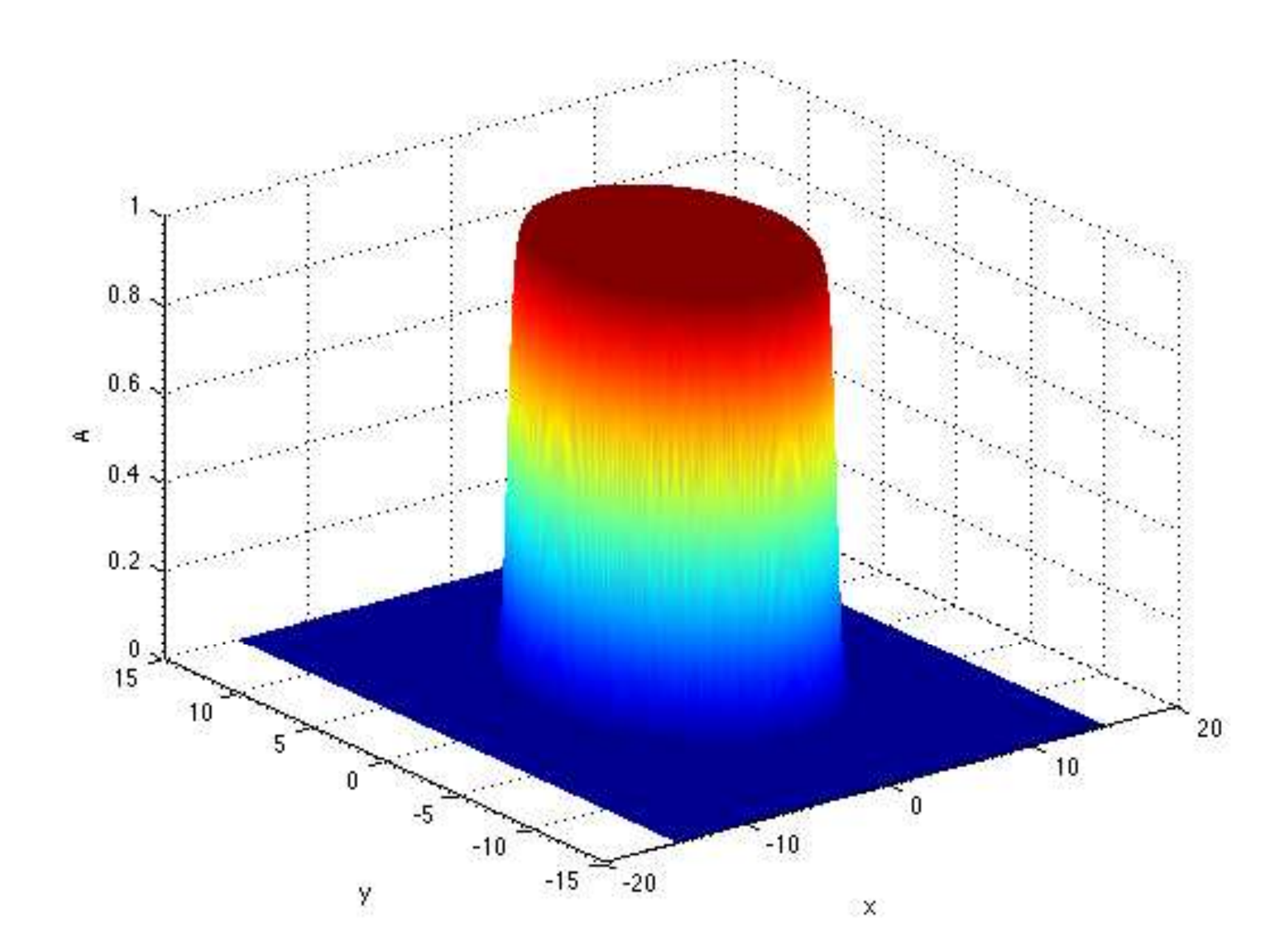} 
		\caption{Initial condition in a 2D dam-break problem.}
	\label{dambreak}
 \end{figure}

We now consider a 2D version of the dam-break problem. 
In the same two-layer model, the fluid heights of both upper and lower layers are assumed to be greater in the central area of a circular dam, which is  released at time $t=0$. 
To avoid numerical instability at the sharp boundaries, we use a smoothed initial condition:
$$\tilde A|_{t=0}=\frac 12 Q \left [\tanh \left (-\nu(x^2+y^2-\tilde{r_0}^2) \right )+1\right ],$$
where  $\tilde{r_0}$ describes the position of the dam, and $\nu$ and $Q$ are suitable positive constants. 
The initial condition is shown in Figure \ref{dambreak} (with $Q=1,~\nu=0.15,$ and $\tilde{r_0}=8$). Note that the wave height parameter $Q$ can be chosen arbitrarily, due to a possible scaling in the problem. 

\subsubsection{Numerical results for surface waves}
\begin{figure}
\centering
	\includegraphics[width=0.55\columnwidth]{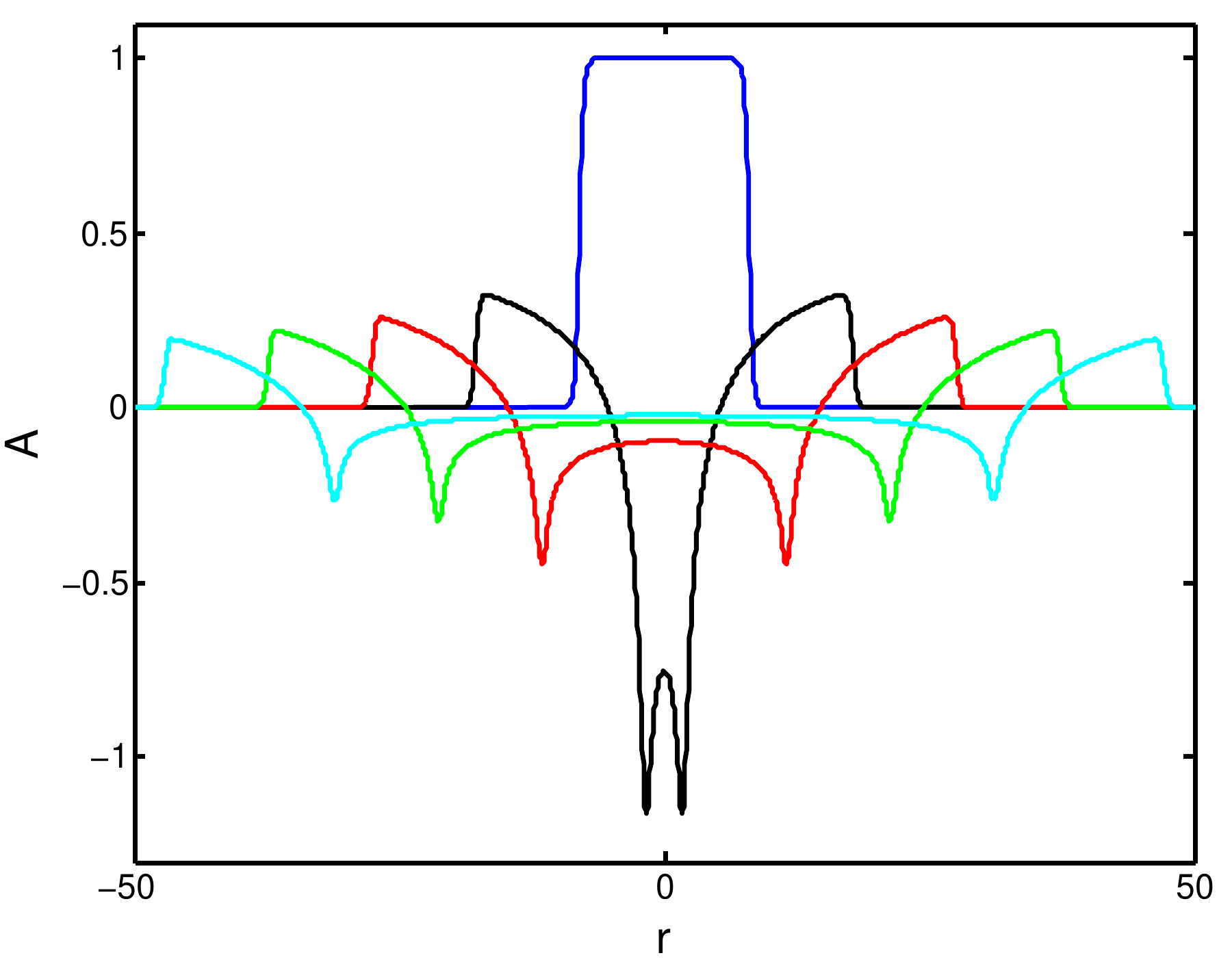} 
		\caption{Linear surface waves for $d=0.6$ and $ Q=1$ at 
		$t=0, 10, 20, 30$ and $40$ (from the `dam-break' initial condition in the centre at $t=0$.
		}
	\label{linearsur06}
 \end{figure}

We choose $d=0.6$ and numerically solve the linear Cauchy problem:
\begin{align*}
\tilde{A}_{tt}-s_+^2(\tilde{A}_{xx}+\tilde{A}_{yy})=0,\\
\tilde{A}|_{t=0}=\frac 12 Q \left [ \tanh \left (-0.15(x^2+y^2-64) \right )+1\right ], \tilde A_t|_{t=0} = 0.
\end{align*}
The cross-section $y=0$ of the linear solution is shown 
in Figure \ref{linearsur06} for $Q=1$. 

Then the numerical solution of the linear problem is used as the initial condition for the derived cKdV equation. Here we let $\varepsilon=0.02$ and impose the initial condition at 
$R = R_0=
0.24$.
The nonlinear Cauchy problem for the surface mode is given by
\begin{align*}
2.1358A_R+3.2015AA_\xi+0.3236A_{\xi\xi\xi}+1.0679 \frac{A}{R}=0,\\
A(\xi,0.24)=\tilde A \left (12,0,\frac{12-\xi}{0.9789}\right ).
\end{align*}

\begin{figure}
\centering
\renewcommand{\thesubfigure}{(\arabic{subfigure})}
\subfigure[$Q=20$]{\includegraphics[width=0.27\textwidth]{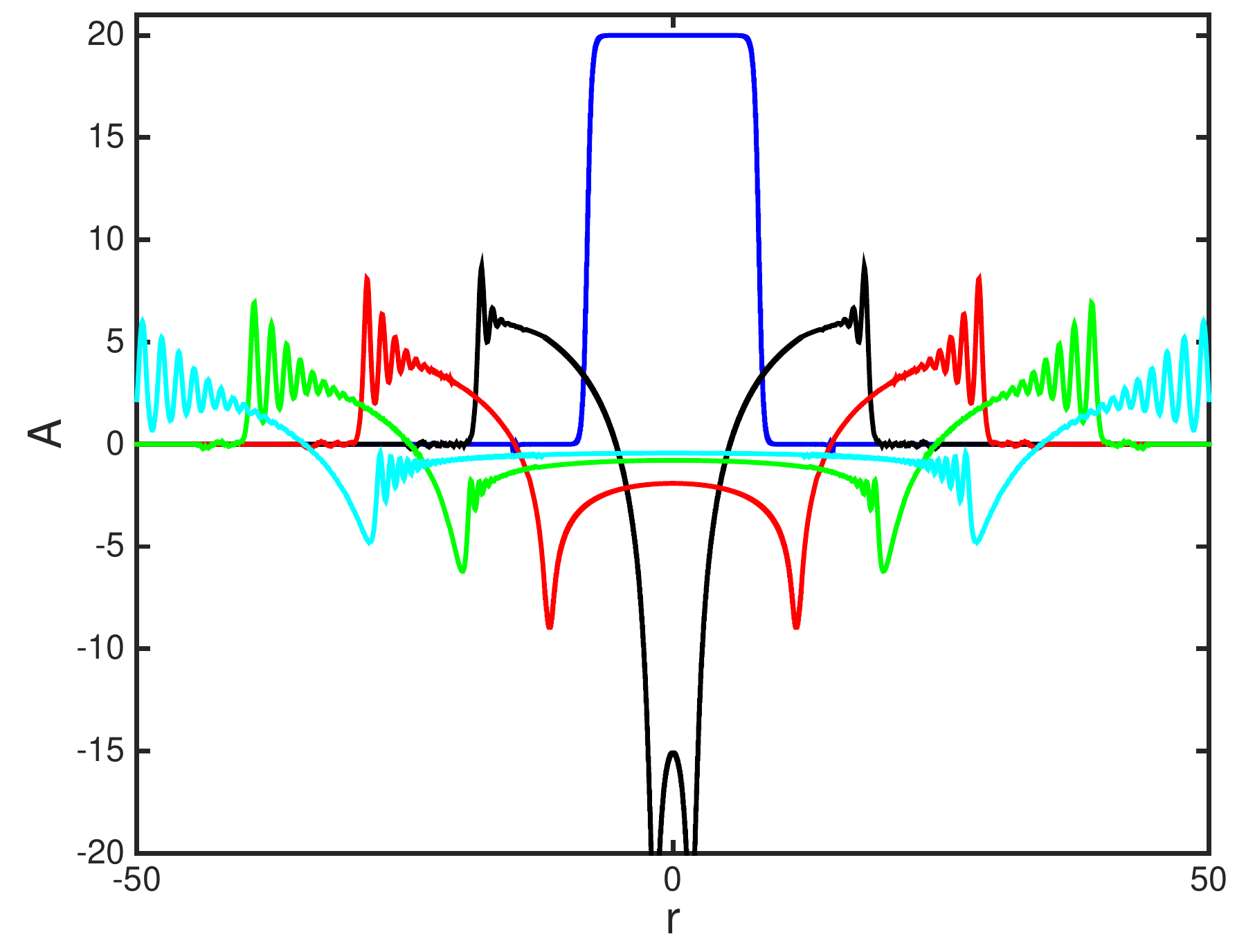}}
\subfigure[$Q=30$]{\includegraphics[width=0.27\textwidth]{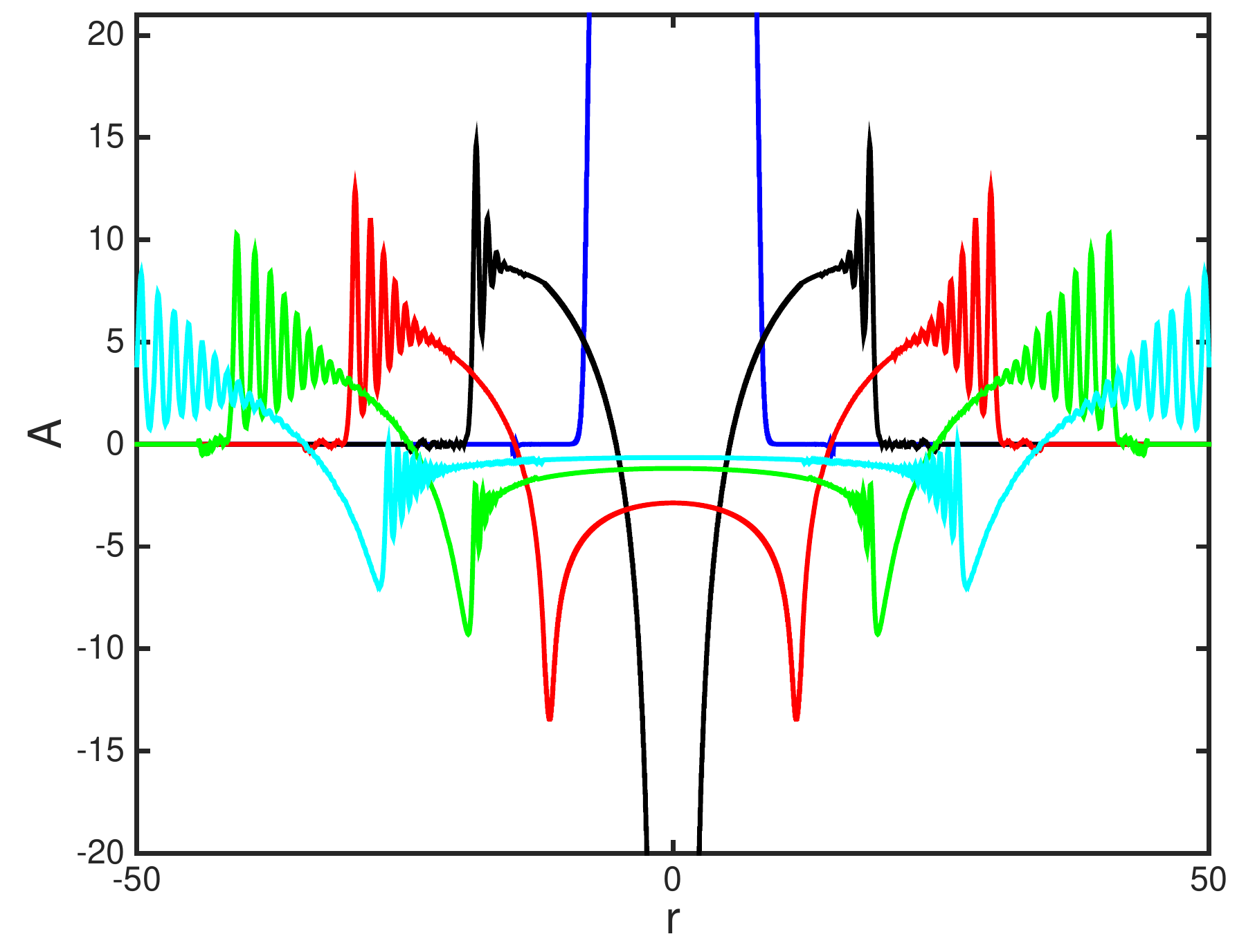}}
\subfigure[$Q=40$]{\includegraphics[width=0.27\textwidth]{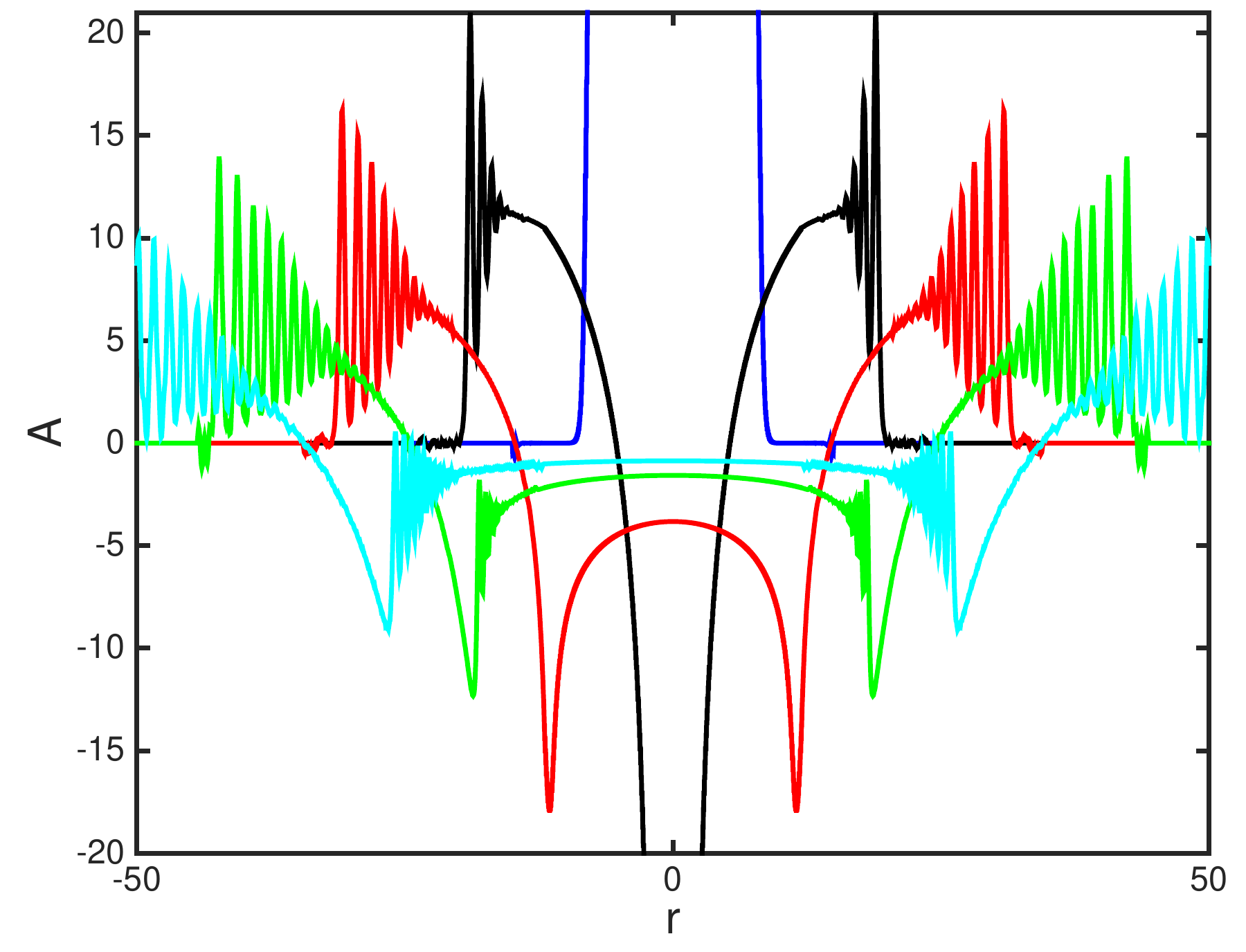}}
\caption{Nonlinear surface waves in the directions $\theta = 0$  and $\theta = \pi$  for $d=0.6$ at
$t=0, 10, 20, 30$ and $40$ (from the `dam-break' initial condition in the centre at $t=0$). 
}
\label{2nssb}
\setcounter{subfigure}{0}
\end{figure}

The cross-section of the numerical solution along the directions $\theta=0$ and $\theta=\pi$ 
for $Q=20, 30$ and $40$ is shown in Figure \ref{2nssb}. 
We see the formation of concentric dispersive shock waves (DSWs), similar to the plane waves described by the KdV equation (see \cite{gurevich1974,hoefer2009} and references therein).   Concentric DSWs have been observed and modelled in Bose-Einstein condensates \cite{hoefer2006,kamchatnov2004}. Whitham's averaging method (see \cite{whitham_book}) became the key analytical tool for the description of such waves (e.g., \cite{smyth1988,el2007,ablowitz2015} and references therein).
The 1D dam-break problem for a two-layer fluid was extensively studied by Esler and Pearce within the framework of the Miyata-Choi-Camassa model \cite{esler2011} (see also \cite{Chumakova}). Relevant experimental observations have been reported in \cite{grue}. 

\begin{figure}
\centering
\renewcommand{\thesubfigure}{(\arabic{subfigure})}
\includegraphics[width=0.27\textwidth]{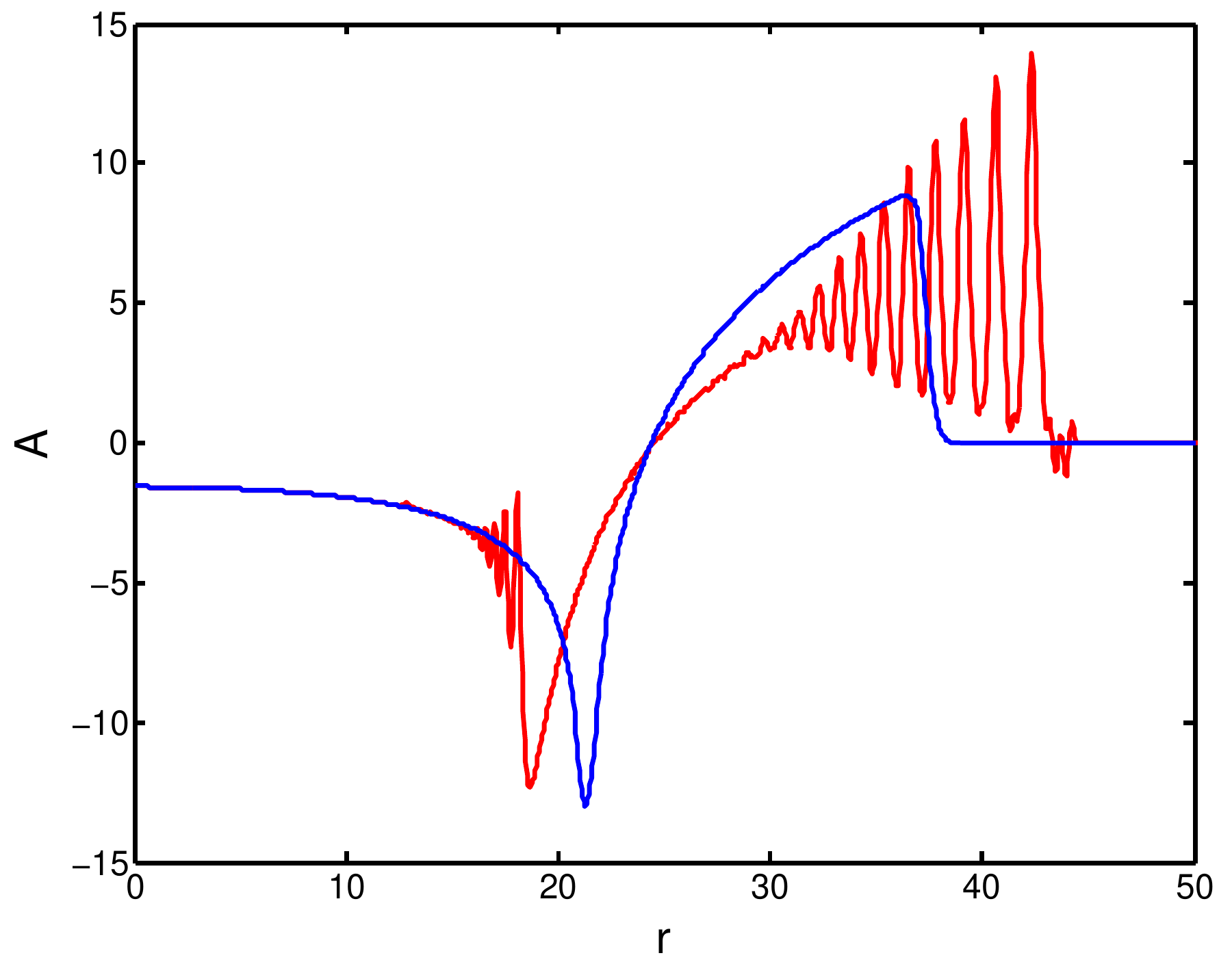}
\caption{The linear (non-oscillatory)
and nonlinear (oscillatory)
surface waves for $Q = 40$ at $t=30$.}
\label{figuresln}
\setcounter{subfigure}{0}
\end{figure}

\begin{figure*}
\centering
	\includegraphics[width=1.6 \columnwidth]{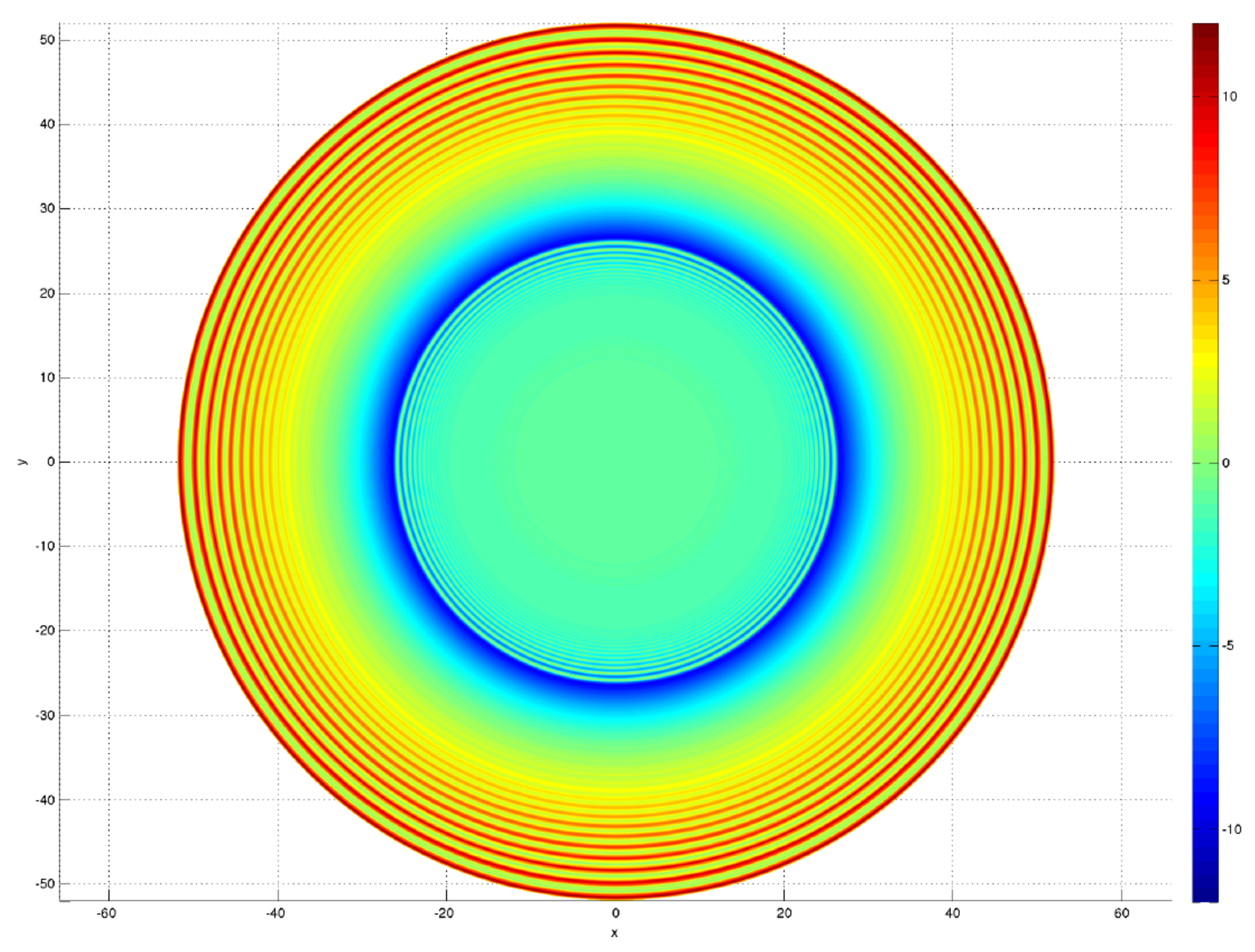} 
		\caption{Nonlinear surface wave for $Q=40$ and $t=40$.
		}
	\label{DSWs}
 \end{figure*}

The linear and the nonlinear solutions are compared in Figure \ref{figuresln} for $Q=40$ at $t=30$. The nonlinear wave propagates faster than the linear wave, and nonlinearity and dispersion, acting together, yield the formation of two concentric DSWs for the initial condition used in this numerical experiment. The surface DSWs for $Q=40$ at $t=40$ are shown in the relief plot in Figure \ref{DSWs}.

\subsubsection{Numerical results for interfacial waves}
\begin{figure}[h]
\centering
	\includegraphics[width=0.55\columnwidth]{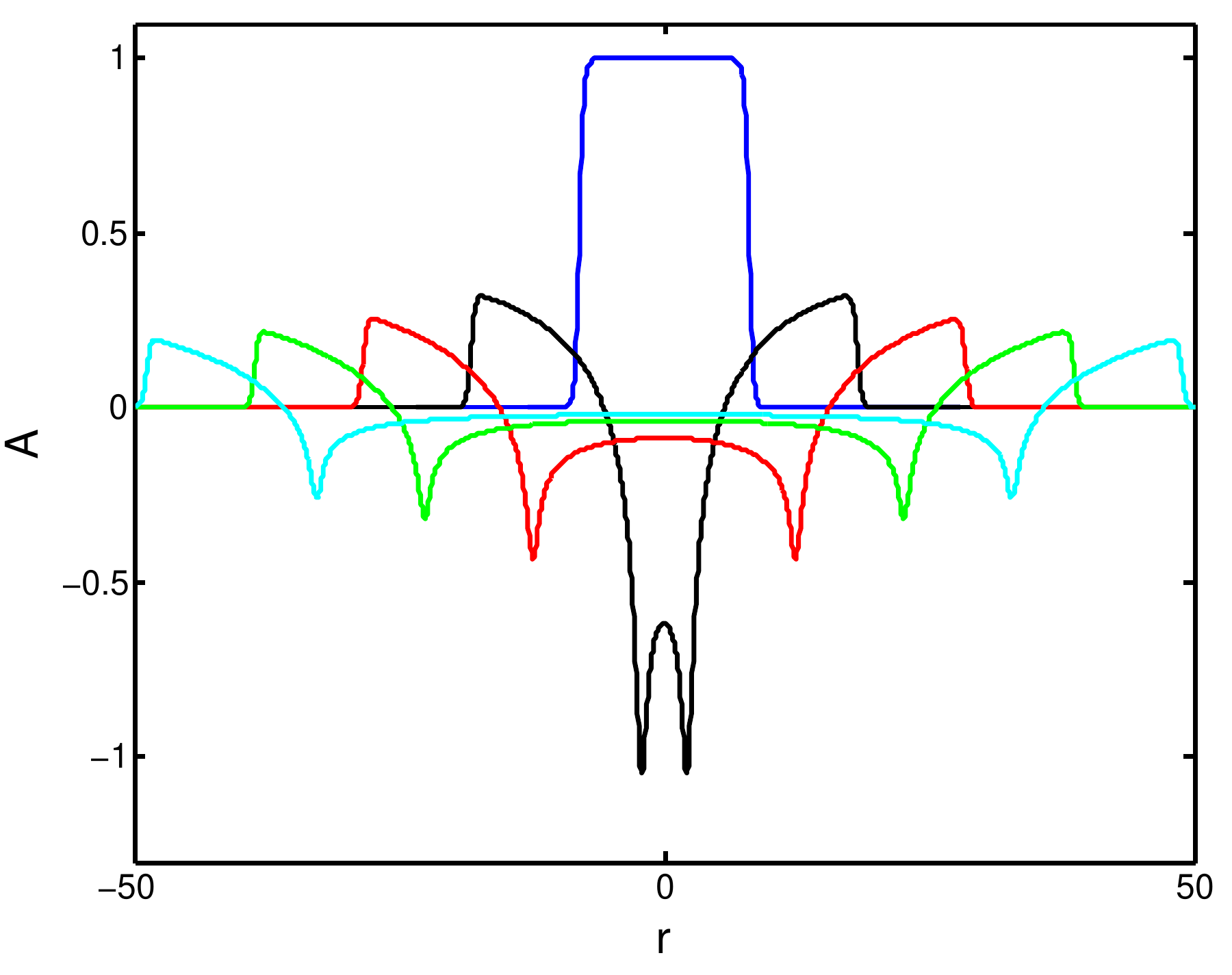} 
		\caption{Linear internal waves 
		for $d=0.6$ 
		at $t=0, 50, 100, 150$ and $200$ (from the `dam-break' initial condition in the centre at $t=0$).
		}
	\label{2lineari}
 \end{figure}

We again choose $d = 0.6$ and numerically solve the linear Cauchy problem
\begin{eqnarray*}
\tilde A_{tt}-s_-^2(\tilde A_{xx}+\tilde A_{yy})=0, \\
\tilde A|_{t=0}=\frac 12 Q \left[\tanh \left(-0.15(x^2+y^2-64)\right)+1 \right], \tilde A_t|_{t=0} = 0,
\end{eqnarray*}
to describe the initial evolution of the waves. The cross-section $y=0$ of the linear solution 
is shown in Figure \ref{2lineari} for $Q=1$  ($Q$ is a scaling factor in the problem).

 \begin{figure}
\centering
\renewcommand{\thesubfigure}{(\arabic{subfigure})}
\subfigure[$Q=20$]{\includegraphics[width=0.27\textwidth]{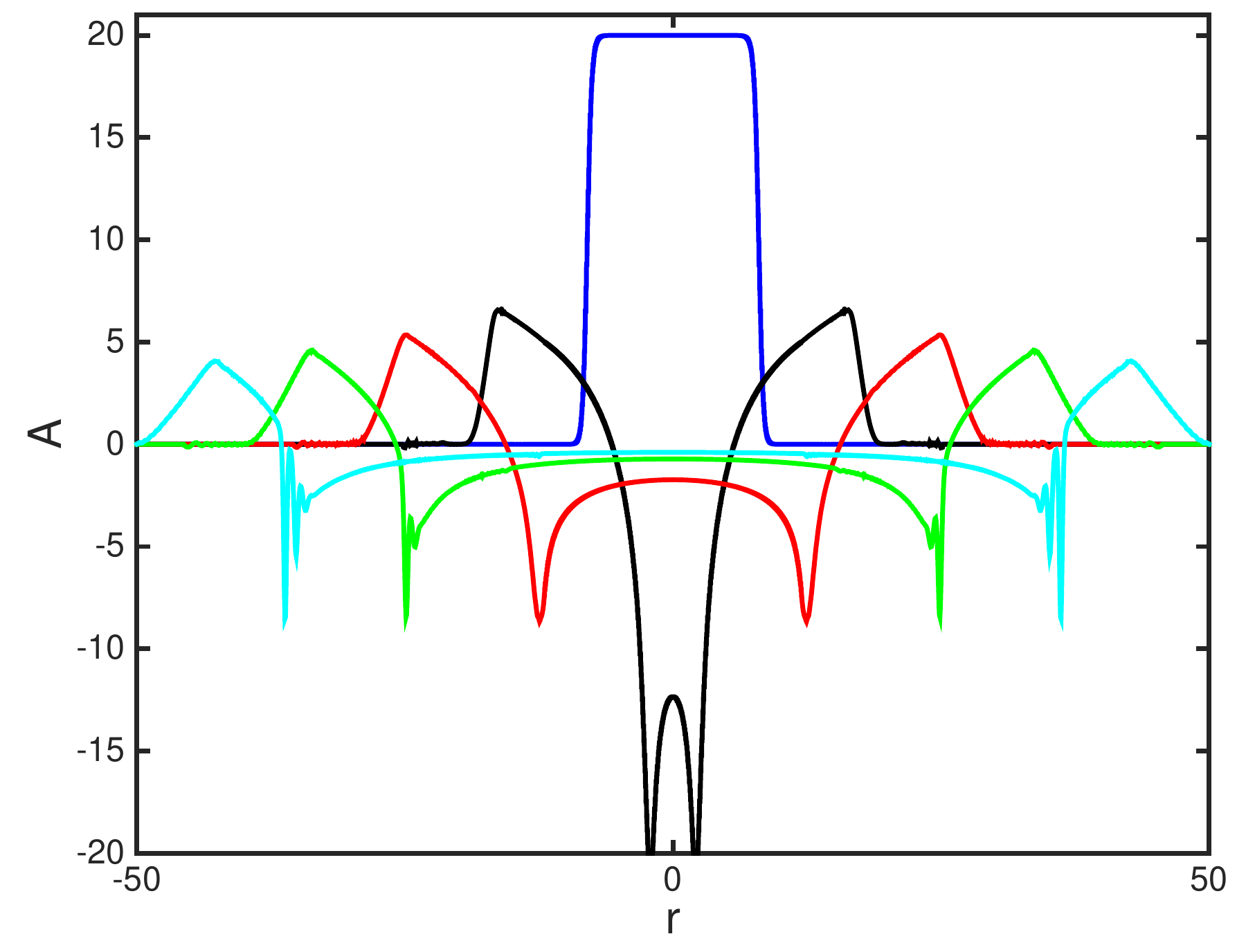}}
\subfigure[$Q=30$]{\includegraphics[width=0.27\textwidth]{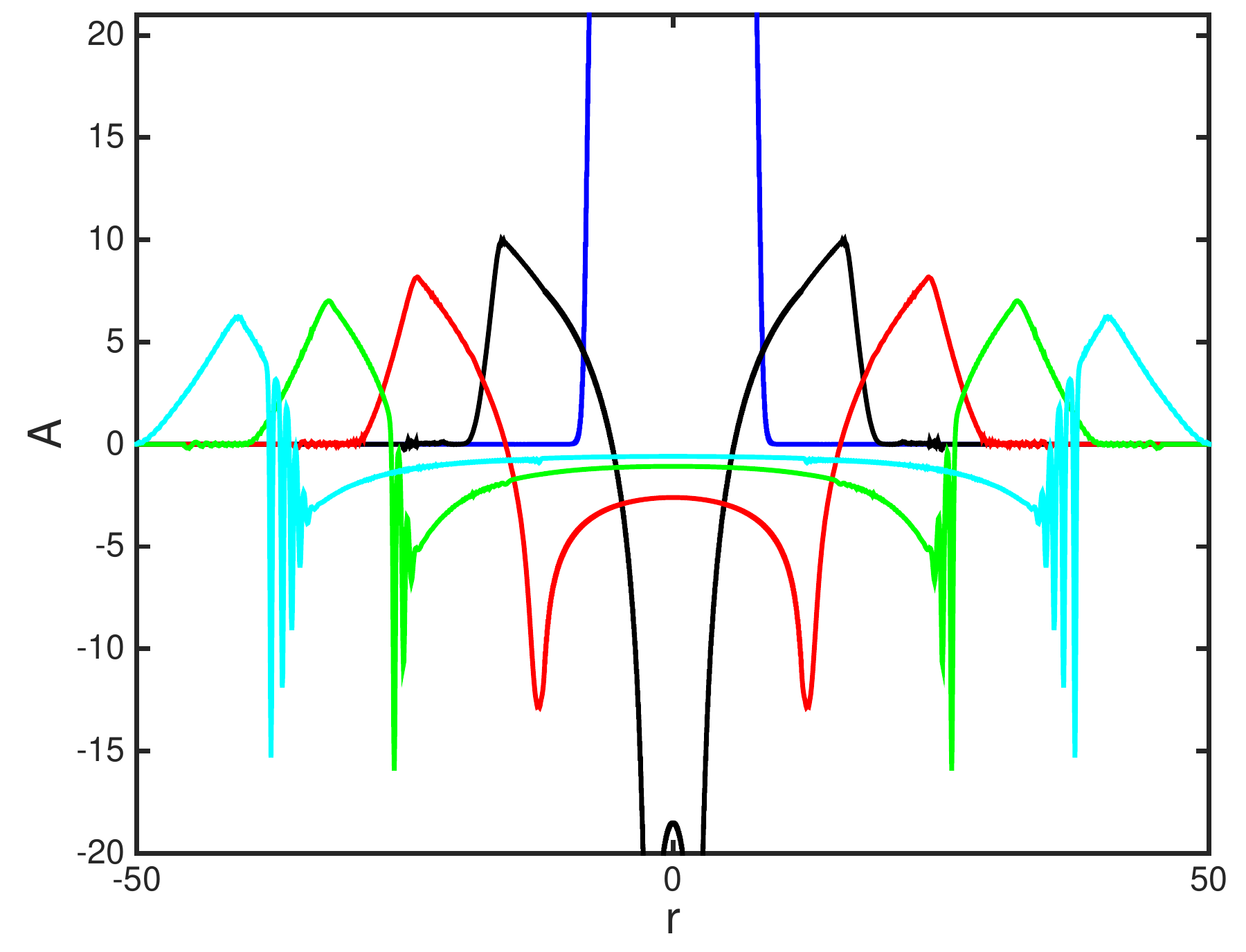}}
\subfigure[$Q=40$]{\includegraphics[width=0.27\textwidth]{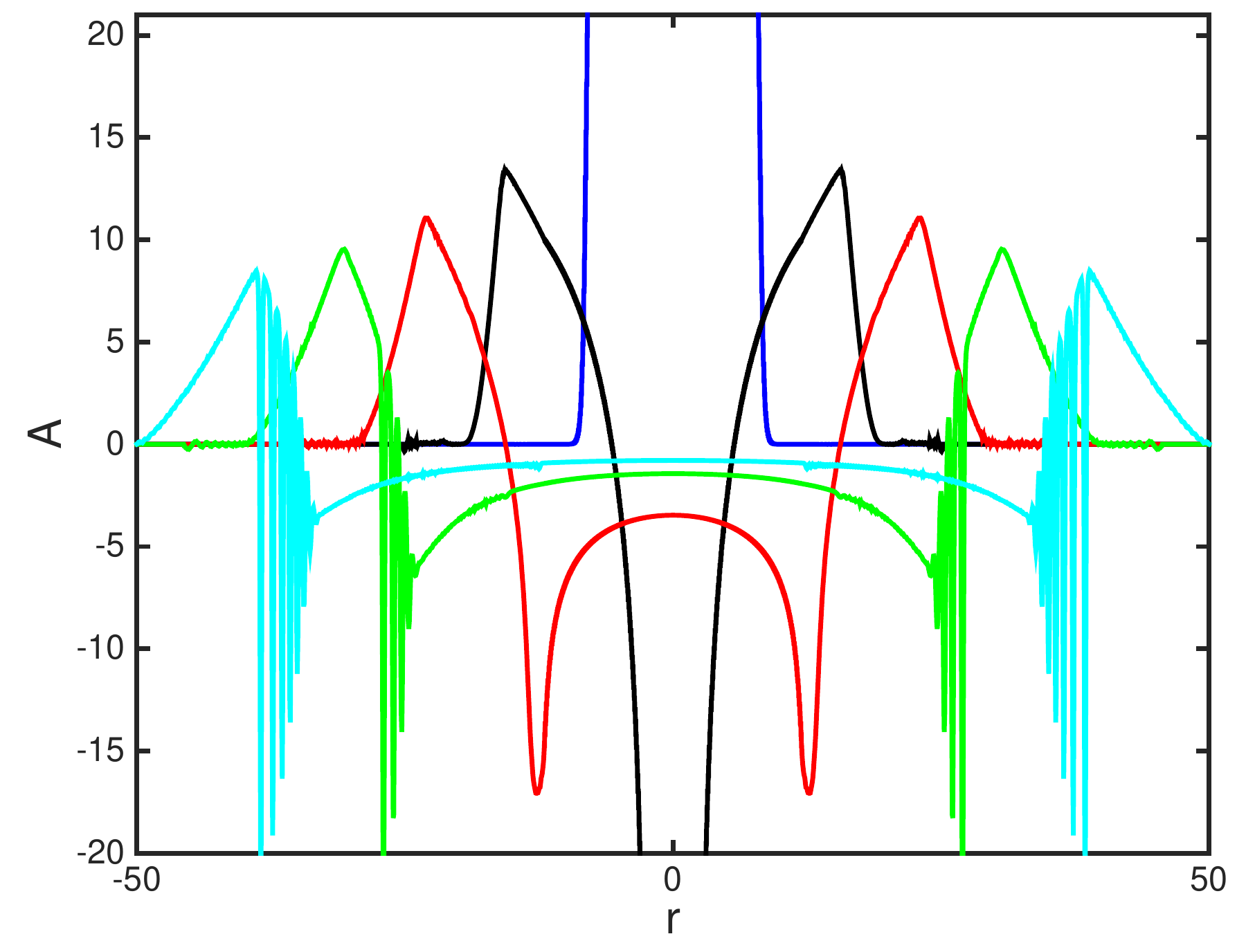}}
\caption{Nonlinear interfacial waves in the directions $\theta = 0$  and $\theta = \pi$  for $d=0.6$ 
at $t=0, 50, 100, 150$ and $200$ (from the `dam-break' initial condition in the centre at $t=0$). 
}
\label{2nis}
\setcounter{subfigure}{0}
\end{figure} 

\begin{figure}
\centering
\renewcommand{\thesubfigure}{(\arabic{subfigure})}
\subfigure[$Q=40$]{\includegraphics[width=0.27\textwidth]{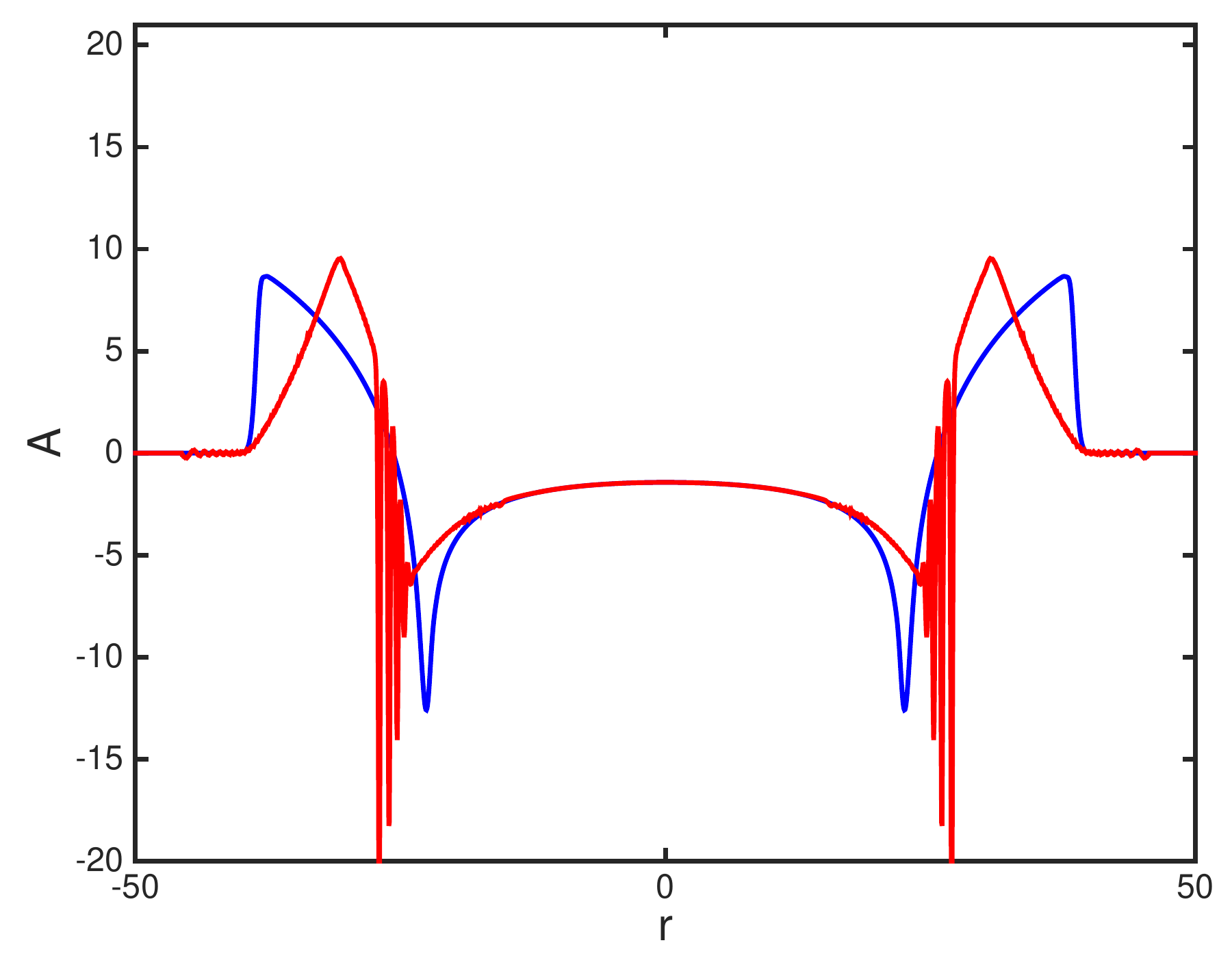}}
\caption{The linear (non-oscillatory)
and nonlinear (oscillatory)
interfacial waves for $Q=40$ at $t=150$. }
\label{figureiln}
\setcounter{subfigure}{0}
\end{figure} 

\begin{figure*}
\centering
	\includegraphics[width=1.6\columnwidth]{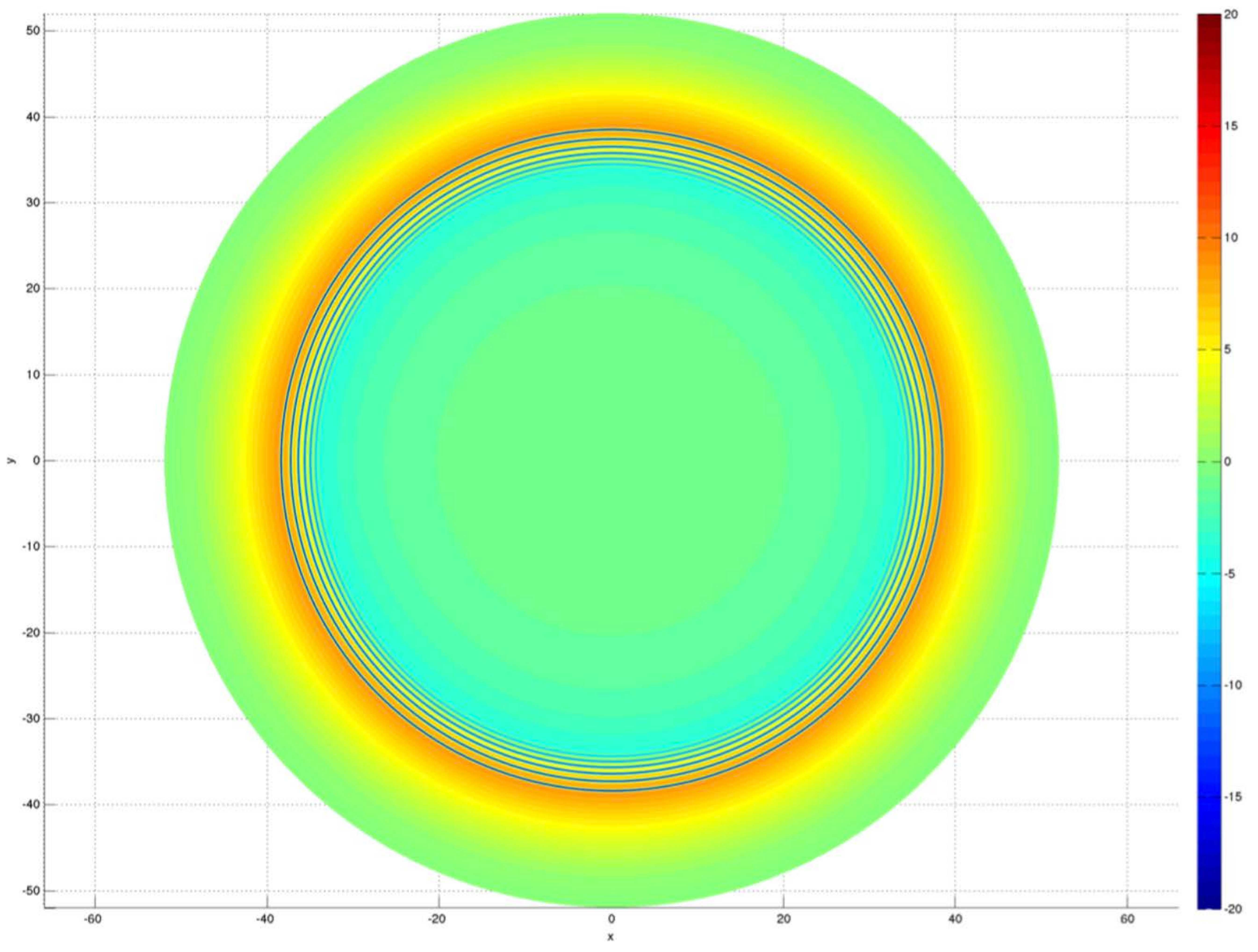} 
		\caption{Nonlinear interfacial wave for $Q=40$ and $t=200$.
		}
	\label{DSWi}
 \end{figure*}

Then, the numerical solution of the linear problem is used as the initial condition for the cKdV equation, with
$\varepsilon=0.02$, 
and $R_0=0.24$. The nonlinear Cauchy problem is given by
\begin{align*}
0.4272A_R-0.6719AA_\xi+0.0150A_{\xi\xi\xi}+0.2136 \frac{A}{R}=0,\\
A(\xi,0.24)=\tilde A \left (12,0,\frac{12-\xi}{0.2043}\right ).
\end{align*}

In Figure \ref{2nis}, the cross-section of the numerical solution is shown along the directions $\theta=0$ and $\theta=\pi$ 
for $Q=20,~30$ and $40$. In Figure \ref{figureiln} the linear and nonlinear solutions are compared for $Q=40$ at $t=150$. The modelling again shows the formation of the concentric DSWs. However, the internal DSWs in Figure \ref{2nis} and Figure \ref{figureiln}  look distinctively different from the surface DSWs shown in Figure \ref{2nssb} and Figure \ref{figuresln}. In particular, there is only one internal DSW formed in the middle range of the relevant linear solution, while there are two surface DSWs formed in the front and back regions of the linear solution, at least for the initial conditions used in these numerical experiments. The internal DSW is shown for $Q=40$ and $t=200$ in the relief plot in Figure \ref{DSWi}.

\section{Ring waves on a shear flow}
In this section, we illustrate the effect of the piecewise-constant shear flow on the ring waves. 
We use a model initial condition, defined by a distorted solution of the 2D linear wave equation \cite{Dobrokhotov}, where the wave amplitude depends on a direction. Then we numerically solve the Cauchy problem for the cKdV-type equation (\ref{cKdV}) with this initial condition. 


\subsection{Model initial condition}

\begin{figure}
\centering
	\includegraphics[width=0.55\columnwidth]{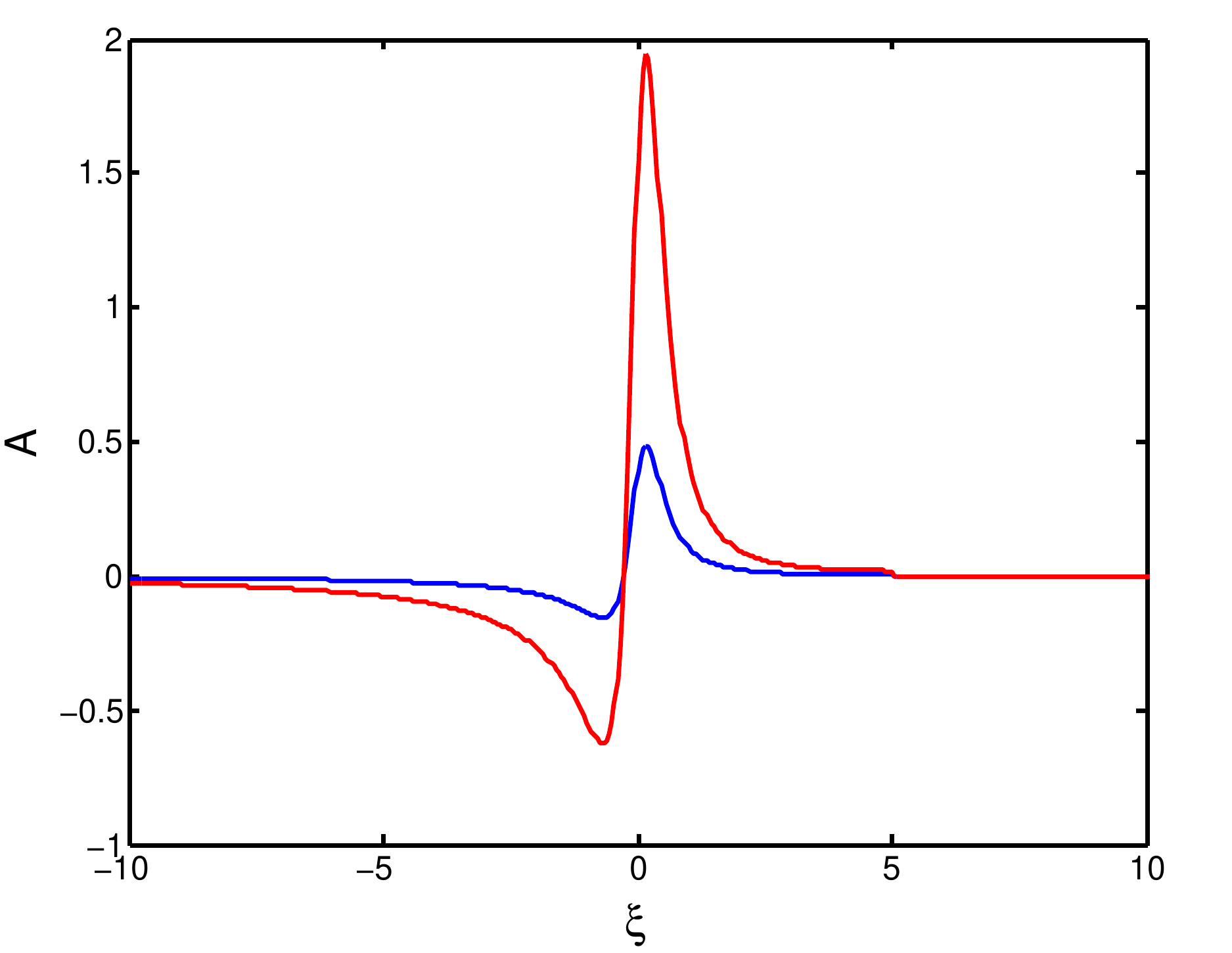} 
		\caption{The initial condition in the directions $\theta=0$ (downstream, 
		lower wave) and $\theta=\pi$ (upstream, 
		higher wave) at $R_0=0.1$. }
	\label{6initial}
 \end{figure}

Recently, Arkhipov et. al. have studied long nonlinear ring waves on the interface of a two-layer fluid with a piecewise-constant shear flow using a coupled system of Bousinesq-type equations \cite{ark2013}  .
In this section, we solve a qualitatively similar problem using the following model initial condition: at $R_0=0.1$ (where $R_0=\varepsilon r_0(\theta) k(\theta)$ with $k(\theta)$ given in Section 3), we define
\begin{eqnarray*}
A(R_0, \xi, \theta)=5 \left(4-\frac{3|\pi-\theta|}{\pi}\right )\qquad \quad\qquad\qquad\\
\qquad \times \  \mbox{Re} \left( \frac {1+2 i(50 R_0-\xi)} {((1+2i(50 R_0-\xi))^2+(100R_0)^2)^{3/2}}\right),
\end{eqnarray*}
which constitutes a distorted analytical solution to the 2D linear wave equation used 
in Section III, where now the amplitude depends on the direction. This model condition is shown in Figure \ref{6initial} for the directions $\theta=0$ and $\pi$, with $R_0=0.1$. Here, the wave height is four times higher in the upstream direction  ($\theta=\pi$) than downstream ($\theta = 0$).   This model initial condition mimics the properties of the waves in \cite{ark2013}: the wave is lower downstream and higher upstream.
This initial condition is discontinuous at $r=0$. However, the initial condition is used at $r=r_0(\theta)>0$ in the time interval $t\in [0,t_1]$. 
We would like to compare the qualitative features of our model problem with the solutions in \cite{ark2013}.

We use the same parameters as before. The densities of the two layers are $\rho_{1}=1$, $\rho_{2}=1.2$ and $\varepsilon=0.02$. Two values of the depth of the lower layer are  $d=0.5$ and $d=0.6$. 

\subsection{Numerical results for interfacial waves}
\begin{figure}
\centering
\renewcommand{\thesubfigure}{(\arabic{subfigure})}
\subfigure[$d=0.5$]{\includegraphics[width=0.27\textwidth]{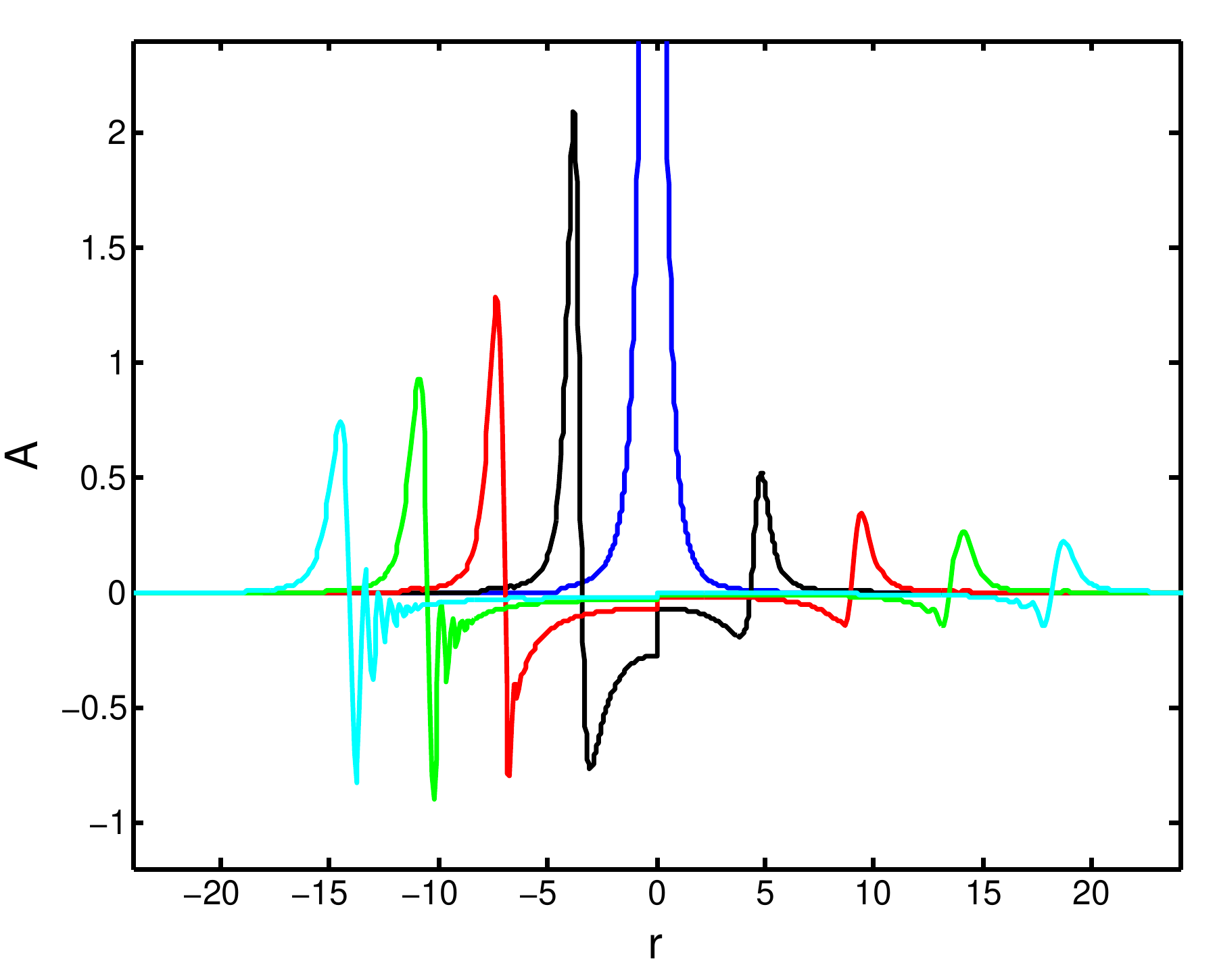}}
\subfigure[$d=0.6$]{\includegraphics[width=0.27\textwidth]{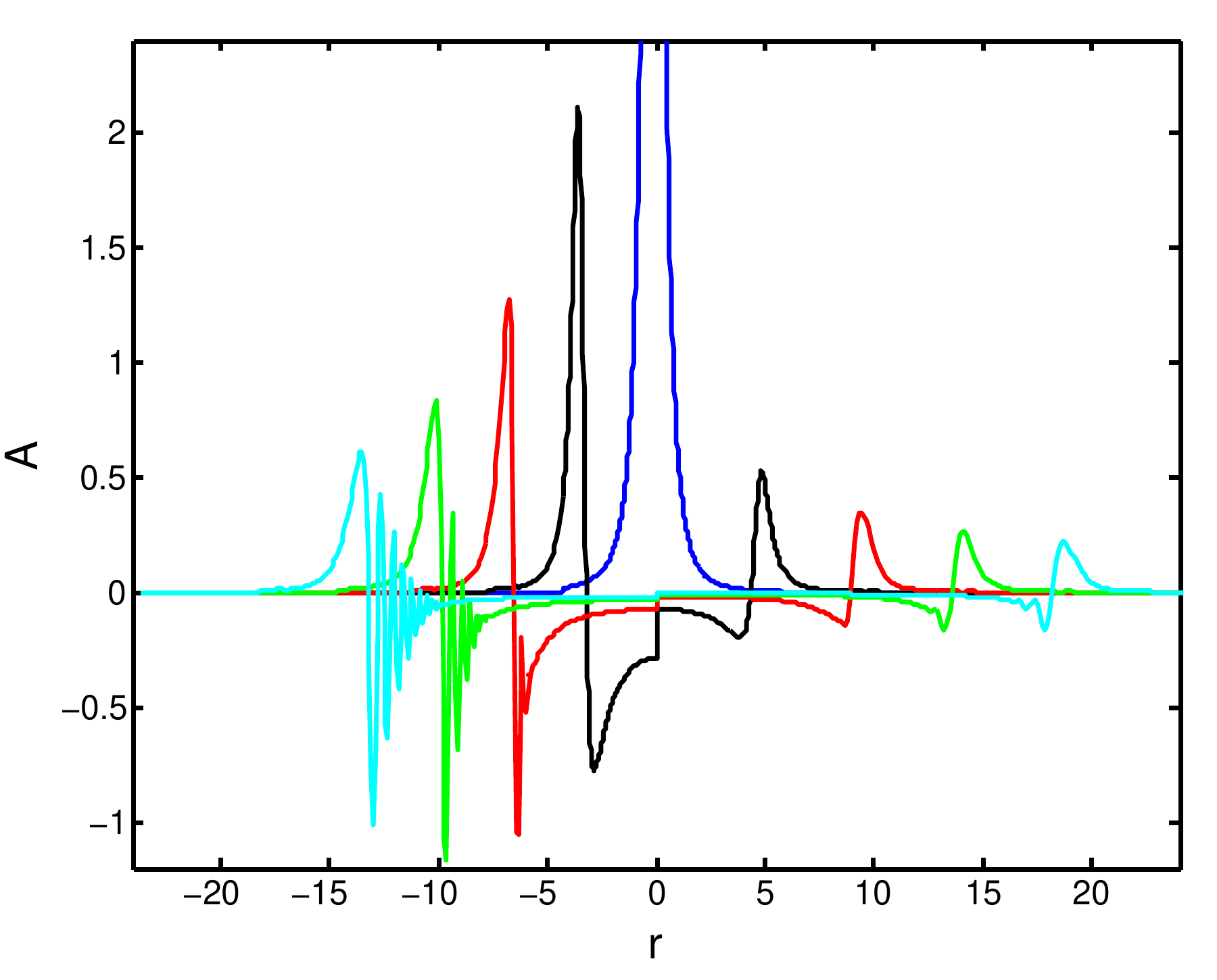}}
\caption{Interfacial waves in the directions $\theta = 0$ (downsteam) and $\theta = \pi$ (upstream) for $U_1-U_2=0.05$
at $t=0, 20, 40, 60$ and $80$ (from a distorted solution for a localised initial condition in the centre at $t=0$).
}
\label{sum61}
\setcounter{subfigure}{0}
\end{figure}

\begin{figure}
\centering
\renewcommand{\thesubfigure}{(\arabic{subfigure})}
\subfigure[$d=0.5$]{\includegraphics[width=0.27\textwidth]{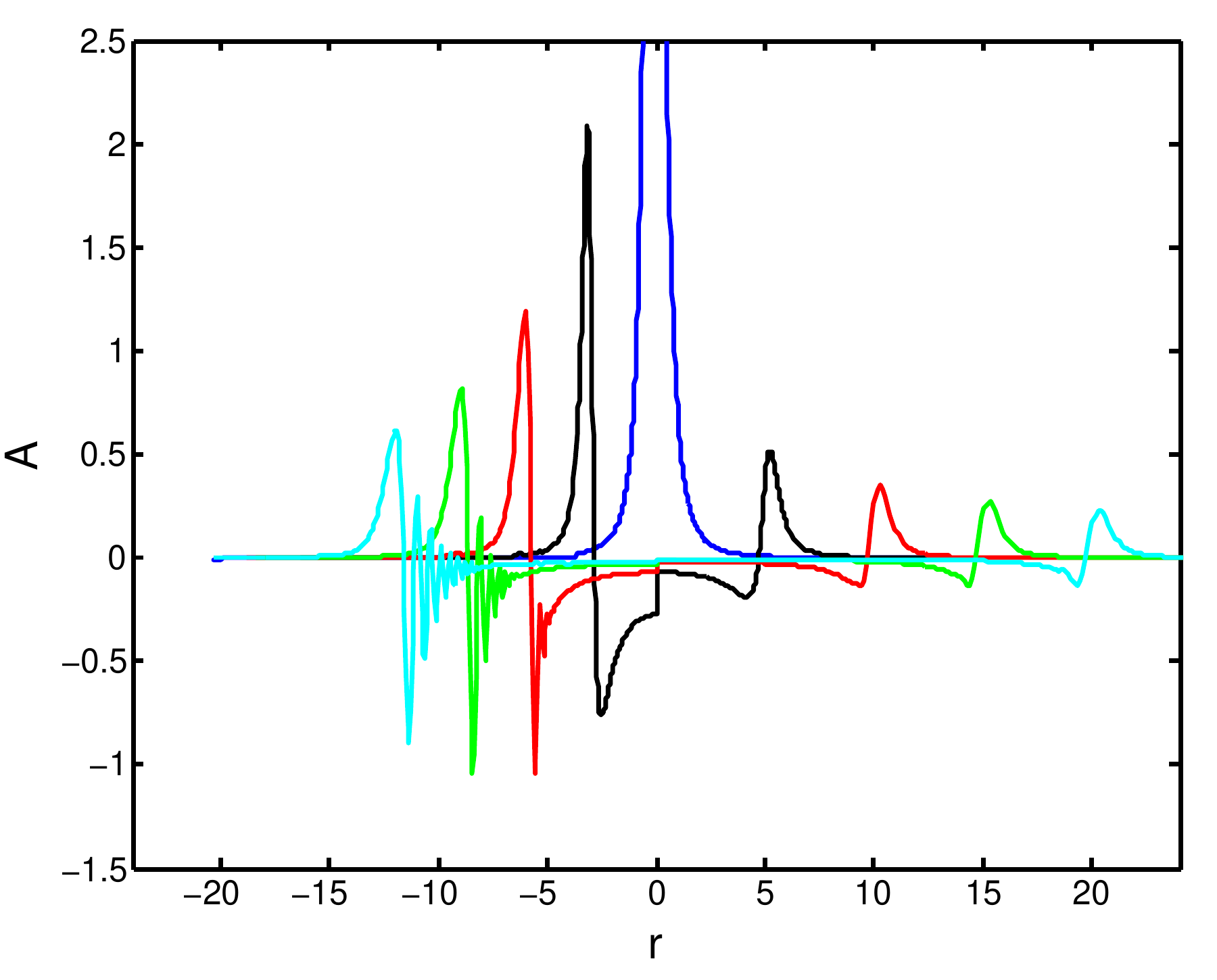}}
\subfigure[$d=0.6$]{\includegraphics[width=0.27\textwidth]{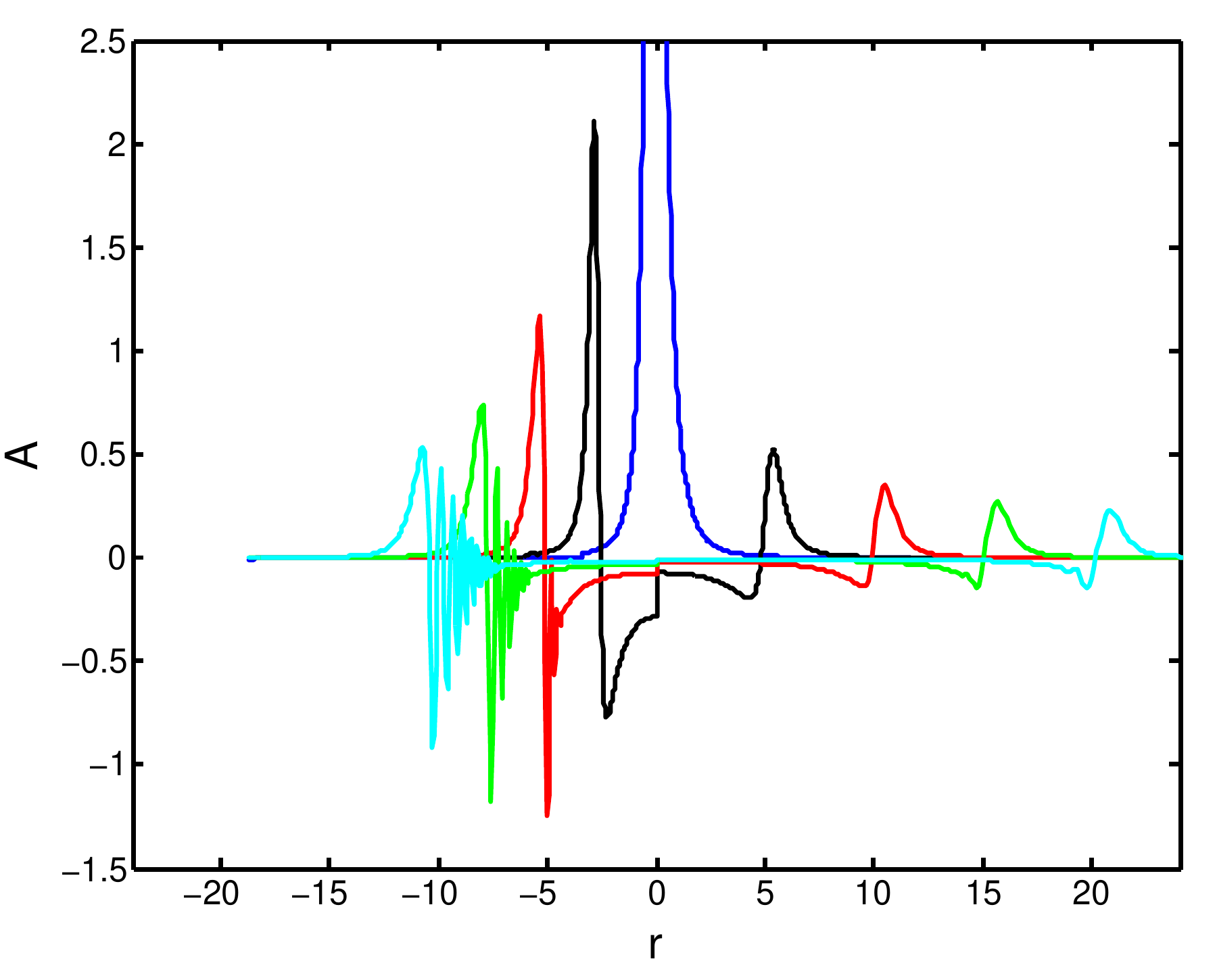}}
\caption{Interfacial waves in the directions $\theta = 0$ (downsteam) and $\theta = \pi$ (upstream) for $U_1-U_2=0.1$
at $t = 0, 20, 40, 60$ and $80$ (from a distorted solution for a localised initial condition in the centre at $t=0$).
}
\label{sum62}
\setcounter{subfigure}{0}
\end{figure}

The $2+1$-dimensional cKdV-type equation (\ref{cKdV}) with variable coefficients is solved numerically using the scheme described in Appendix C. 
We plot the cross-section of the solution in the downstream and upstream directions for $U_1-U_2=0.05$ (Figure \ref{sum61}) and $U_1-U_2=0.1$ (Figure \ref{sum62}). 

\begin{figure}
 \centering
		\includegraphics[width=0.55\columnwidth]{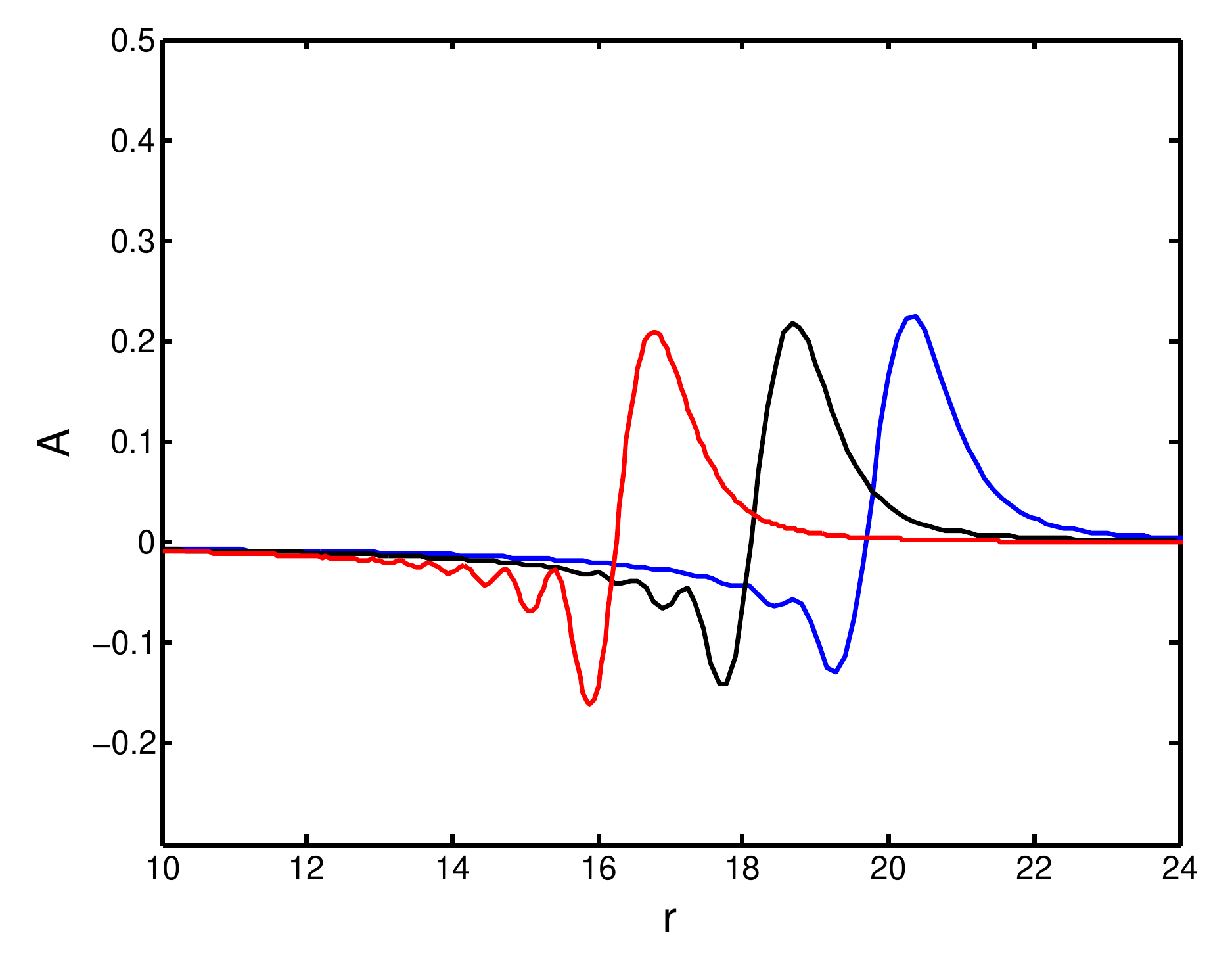} 
		\caption{Interfacial waves in the direction $\theta = 0$ (downsteam)  for $U_1-U_2=0$ (slowest wave), 
		$0.05$ (wave in the middle)
		and $0.1$ (fastest wave)
		at $t=80$. }
	\label{sumsheari}
 \end{figure}

\begin{figure}
 \centering
		\includegraphics[width=0.55\columnwidth]{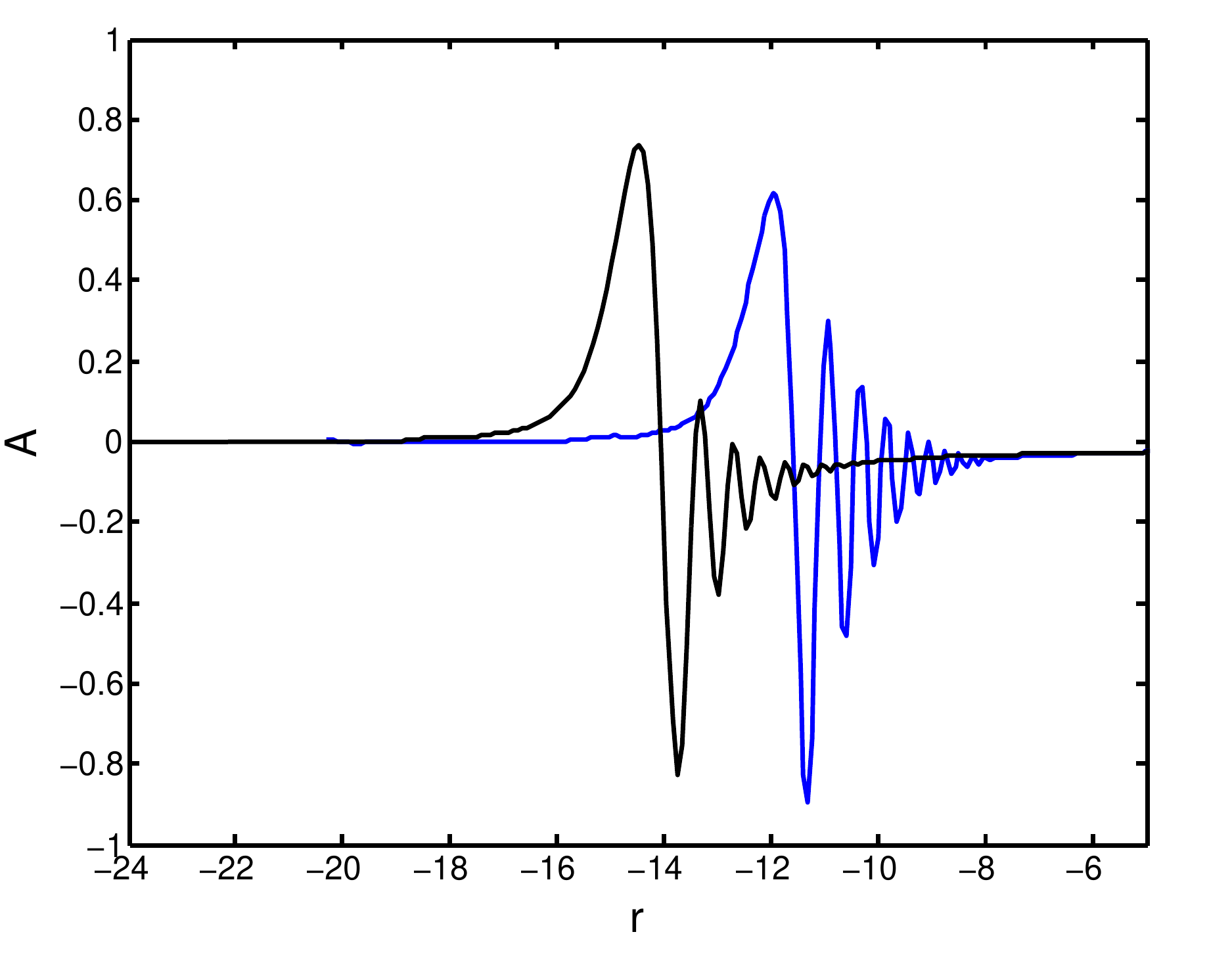} 
		\caption{Interfacial waves in the direction $\theta = \pi$ (upsteam)  for $U_1-U_2=0.05$ (faster wave)
		and $0.1$ (slower wave)
		at $t=80$. }
	\label{sumsheariu}
 \end{figure}

The modelling shows  that the rate of decrease of the height of interfacial waves with the distance from the origin is greater in the upstream direction, 
which agrees with the behaviour of solutions in \cite{ark2013}.
We  plot the interfacial waves in the downstream direction for $d=0.5$ at $t=80$, with $U_1-U_2=0,~ 0.05$, and $0.1$ in Figure \ref{sumsheari}, and in the upstream direction  
in Figure \ref{sumsheariu}.  The shear flow increases the wave speed downstream and decreases the wave speed upstream.
This feature agrees  with the effect of the squeezing of the interfacial wavefronts in the direction of the shear flow  described in \citep{kx14}.  We also note that, with the increase of the strength of the shear flow,
  the rate of the change of the wave height decreases downstream and increases upstream.

\begin{figure}
\centering
\renewcommand{\thesubfigure}{(\arabic{subfigure})}
\subfigure[$d=0.5$]{\includegraphics[width=0.27\textwidth]{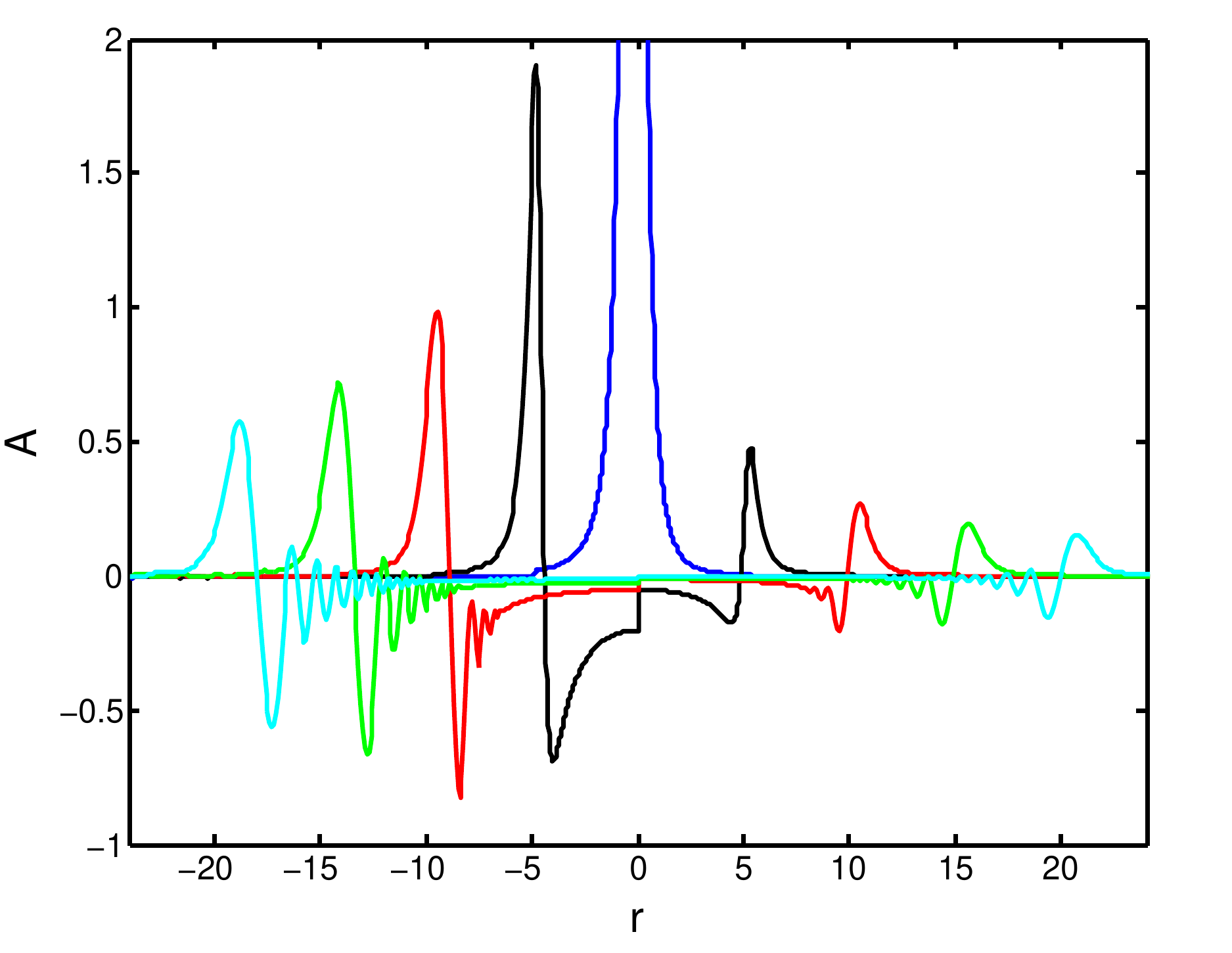}}
\subfigure[$d=0.6$]{\includegraphics[width=0.27\textwidth]{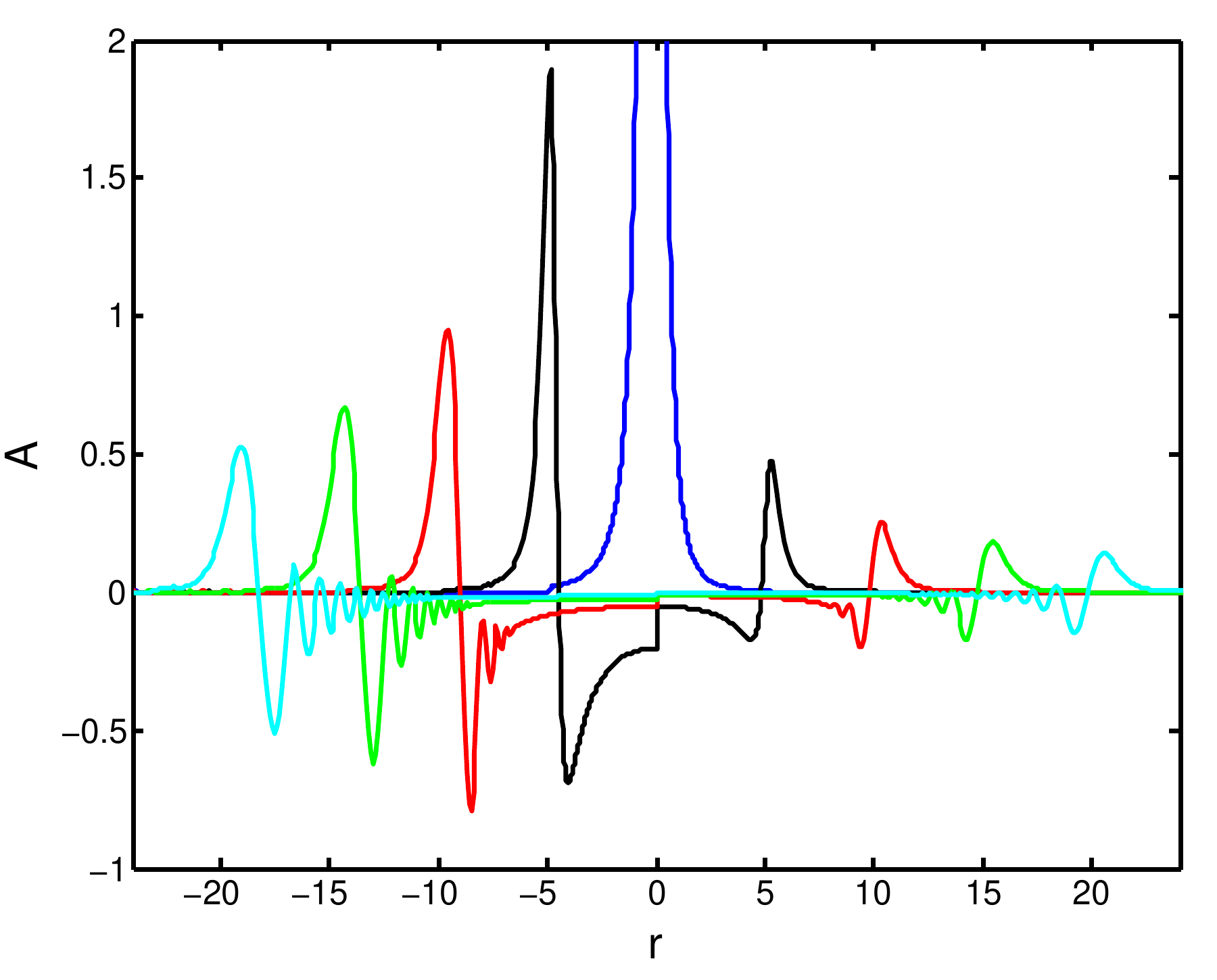}}
\caption{Surface waves in the directions $\theta = 0$ (downsteam) and $\theta = \pi$ (upstream) for $U_1-U_2=0.1$
at $t=0, 5, 10, 15$ and $20$ (from a distorted solution for a localised initial condition in the centre at $t=0$).
}
\label{sum63}
\setcounter{subfigure}{0}
\end{figure}

 \begin{figure}
 \centering
		\includegraphics[width=0.55\columnwidth]{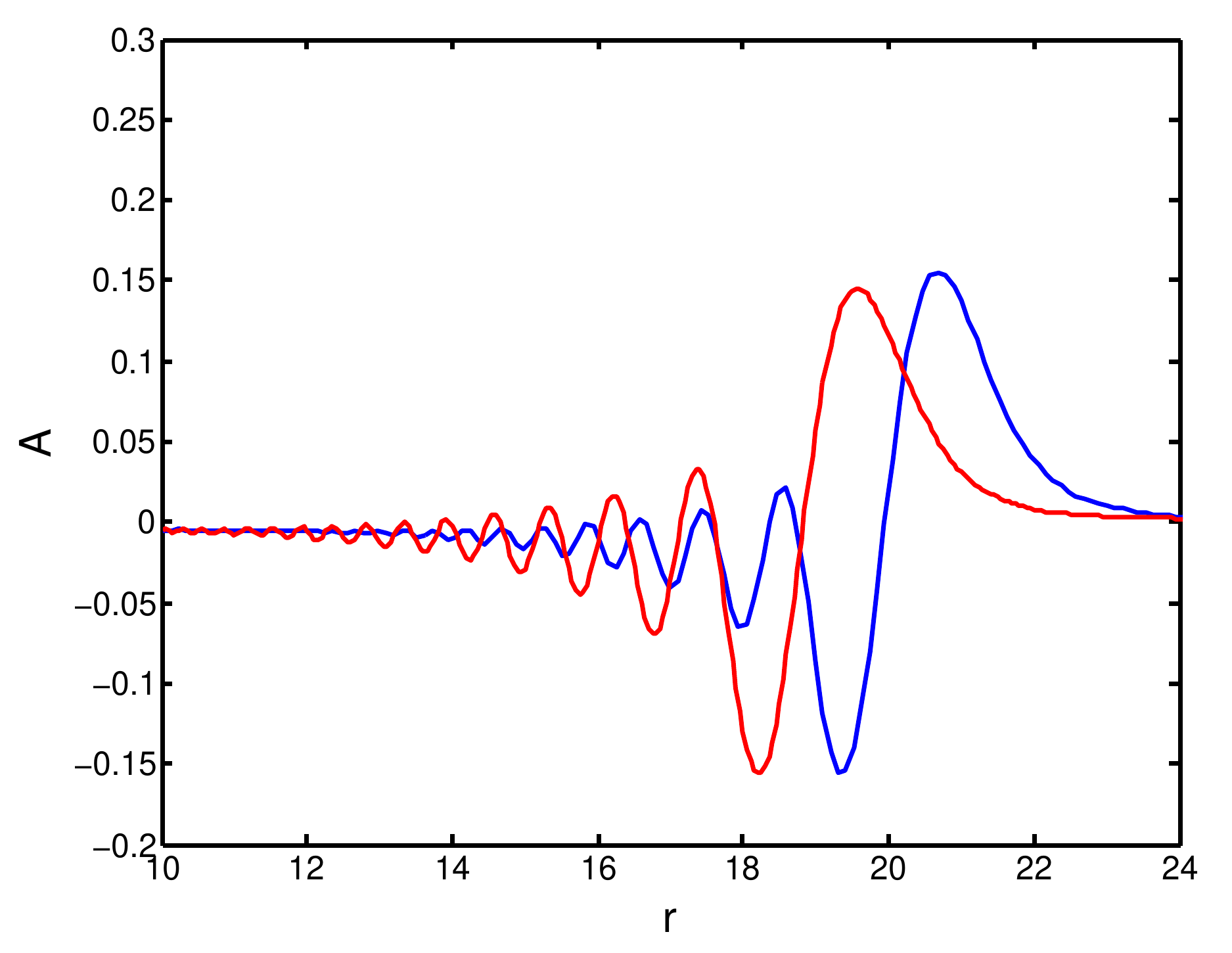} 
		\caption{Surface waves in the directions $\theta = 0$ (downsteam) for  $U_1-U_2=0$ (slower wave)
		and $0.1$ (faster wave)
		at $t=20$. }
	\label{sumshears}
 \end{figure}

\subsection{Numerical results for surface waves}
The cross-section of the numerical solution is shown  in the downstream and upstream directions for $U_1-U_2=0.1$ in Figure \ref{sum63}. Surface waves in the downstream direction  are shown for $t=20$ and $d=0.5$, with  $U_1-U_2=0$ and $0.1$ in Figure \ref{sumshears}.

The wave height also decreases faster upstream 
than downstream. 
 Qualitatively, the shear flow has similar effect on the surface wave height as on the interfacial wave height. Quantitavely, the effect of the weak shear flow on the surface waves is weaker than its effect on the interfacial waves (see Figure \ref{sumsheari}) since the speed of the flow  $U_1-U_2=0.1$ is much smaller than the surface wave speed. However, the important difference is that the shear flow elongates the wavefronts of the surface waves in the direction of the shear flow, while squeezing the wavefronts of the interfacial waves (see \cite{kx14} for details).
 
 We note that a  useful equation was recently derived in \cite{arkhipov2015} in order to describe  axisymmetric surface waves. Unlike the cKdV-type models, this equation allows for initial conditions to be imposed at $r=0$.  However, it does not account for stratification and shear flow.
 
 We also note that the linear solution constructed in \cite{ellingsen14, li} provides the necessary initial condition for the surface ring wave on a shear current of uniform vorticity.

\section{Conclusion}

In this paper we modelled the propagation of surface and interfacial ring waves in a two-layer fluid, using the recently derived 2+1-dimensional cKdV-type equation \cite{kx14}. The numerical finite-difference scheme used in the paper is an extension of the unconditionally stable implicit numerical scheme suggested by Feng and Mitsui in \cite{Feng98}. We considered three particular problems: concentric waves generated from a localised condition and a 2D version of the dam-break problem, as well as more complicated asymmetric ring waves in the presence of a piecewise-constant shear flow. The modelling has shown that the formation of 2D dispersive shock waves and oscillatory wave trains is a typical scenario for the cases under study. Small changes in physical parameters (e.g., the relative depth of the layers) can result in significant changes in the coefficients of the derived equation, and, as a consequence, in significant differences in wave profiles. Weak shear flow has greater effect on interfacial waves, resulting in a number of qualitative and quantitative changes, described in the paper.







\section{Acknowledgments}

We thank G.A.  El, P.A.  Milewski, L.A. Ostrovsky and Yu.A. Stepanyants for references and useful discussions.

\appendix

\section{Coefficients of the cKdV-type equation}

In this Appendix we list the coefficients of the derived $2+1$-dimensional amplitude equation (\ref{cKdV}) for both surface and interfacial ring waves in the two-layer case. 

For the surface waves, we normalise $\phi$ by setting $\phi=1$ at $z=1$. The parameter $\Lambda$ in the modal function (\ref{phi}) is given by
$$\Lambda_s=\frac{k^2+k'^2}{F_1^2}.$$ 
Substituting the modal function into the formulae (\ref{c1}) - (\ref{c5}), we obtain the coefficients in the form
{\small
\begin{align*}
&\mu_1=\frac{2s(k^2+k'^2)^2}{F_1^4} \left((1-d)\rho_1F_1+\frac{\rho_2F_2}{d} \left (\frac{F_1^2}{k^2+k'^2}+d-1\right )^2\right),\\
&\mu_2=-\frac{3(k^2+k'^2)^3}{F_1^6}\left( (1-d)\rho_1F_1^2+\frac{\rho_2F_2^2}{d^2}\left (\frac{F_1^2}{k^2+k'^2}+d-1\right )^3\right),
\\
&\mu_3= -\frac{(k^2+k'^2)^3}{3F_1^4}\bigg(\rho_1F_1^2 \left (\frac{F_1^6}{(k^2+k'^2)^3}- \left (\frac{F_1^2}{k^2+k'^2}+d-1 \right )^3 \right )+\\
&\qquad\qquad\qquad\qquad\qquad\qquad \rho_2F_2^2d \left (\frac{F_1^2}{k^2+k'^2}+d-1 \right )^2\bigg),\\
&\mu_4=-\frac{(1-d)\rho_1k(k+k'')(k^2+k'^2)}{F_1^4}\bigg(F_1^2+4k'F_1(U_1-U_2)\sin\theta\\
&\qquad \qquad\qquad\qquad\qquad +3(k^2+k'^2)(U_1-U_2)^2\sin^2\theta\bigg)\\
& -\frac{\rho_2(k+k'')kF_2^2}{dF_1^4} \left (\frac{F_1^2}{k^2+k'^2}+d-1 \right )\bigg((k^2-3k'^2) \left (\frac{F_1^2}{k^2+k'^2}+d-1\right )+\\
&\qquad\qquad\qquad\qquad\frac{4(d-1)k'(k'F_1+(U_1-U_2)(k^2+k'^2)\sin\theta)}{F_1}\bigg),
\end{align*}
\begin{align*}
&\mu_5=-\frac{2k(k^2+k'^2)}{F_1^4}\bigg((1-d)\rho_1F_1(k'F_1+(U_1-U_2)(k^2+k'^2)\sin\theta)\\
&\qquad\qquad\qquad\qquad+\frac{\rho_2k'F_2^2}{d}\left (\frac{F_1^2}{k^2+k'^2}+d-1\right )^2\bigg),
\end{align*} 
}
where 
\begin{eqnarray*}
F_1&=&-s+(U_1-U_2)(k\cos\theta-k'\sin\theta),\\
F_2&=&-s,\\
s^2&=&\frac{1+\sqrt{(2d-1)^2+4\rho_1/\rho_2 d(1-d)}}{2},
\end{eqnarray*}
and the function $k(\theta)$ is defined by the formula (\ref{solutionk}).

For the interfacial waves, we normalise $\phi$ by setting $\phi=1$ at $z=d$. The parameter $\Lambda$ in the modal function (\ref{phi}) is  given by
$$\Lambda_i=\frac{k^2+k'^2}{F_1^2+(d-1)(k^2+k'^2)}.$$ 
Substituting the modal function into the formulae (\ref{c1}) - (\ref{c5}), we obtain the coefficients in the form 

{\small
\begin{align*}
&\mu_1=2s\left( \frac{(1-d)\rho_1F_1(k^2+k'^2)^2}{(F_1^2+(d-1)(k^2+k'^2))^2}+\frac{\rho_2F_2}{d}\right),\\
&\mu_2=-3\left(\frac{(1-d)\rho_1F_1^2(k^2+k'^2)^3}{(F_1^2+(d-1)(k^2+k'^2))^3}+\frac{\rho_2F_2^2}{d^2}\right),\\
&\mu_3=-\frac{\rho_1F_1^2}{3(F_1^2+(d-1)(k^2+k'^2))^2}\left(F_1^6-(F_1^2+(d-1)(k^2+k'^2))^3\right)\\
&\qquad\qquad\qquad\qquad\qquad\qquad -\frac 13 d \rho_2F_2^2(k^2+k'^2),\\
&\mu_4=-\bigg((k^2-3k'^2)F_1^2-4k'(k^2+k'^2)F_1(U_1-U_2)\sin\theta\\
&\qquad\qquad -(U_1-U_2)^2(k^2+k'^2)^2\sin^2\theta\bigg)\frac{(1-d)\rho_1k(k+k'')}{(F_1^2+(d-1)(k^2+k'^2))^2}\\
&\qquad -\frac{4(1-d)\rho_1k(k+k'')F_1^2}{(F_1^2+(d-1)(k^2+k'^2))^3}(k'F_1+(k^2+k'^2)(U_1-U_2)\sin\theta)^2\\
&\qquad \qquad \qquad \qquad -\frac{\rho_2k(k+k'')(k^2-3k'^2)F_2^2}{d(k^2+k'^2)^2},\\
&\mu_5= -\frac{2(1-d)\rho_1F_1 k(k^2+k'^2)}{(F_1^2+(d-1)(k^2+k'^2))^2}(k'F_1+(U_1-U_2)(k^2+k'^2)\sin\theta)\\
&\qquad \qquad \qquad \qquad \qquad \qquad -\frac{2kk'\rho_2F_2^2}{d(k^2+k'^2)},
\end{align*}
}
where 
\begin{eqnarray*}
F_1&=&-s+(U_1-U_2)(k\cos\theta-k'\sin\theta),\\
F_2&=&-s,\\
s^2&=&\frac{1-\sqrt{(2d-1)^2+4\rho_1/\rho_2 d(1-d)}}{2},
\end{eqnarray*}
and the function $k(\theta)$ is defined by the formula (\ref{solutionk}).


\section{Linear waves in a two-layer fluid}
This is a brief overview of the derivation of the 2D linear wave equations for the surface and interfacial modes in the absence of a shear flow, which follows the approach in \cite{leblond}.

In non-dimensional coordinates, the basic density of the fluid is 
$$ \rho_0= \left\{ 
 \begin{array}{ll}
     \rho_{0(1)}& \mbox{if} \quad z \in  (d,1),\\
     \rho_{0(2)}& \mbox{if} \quad z  \in (0,d),
   \end{array} \right.$$
where $\rho_{0(1)}$ and $\rho_{0(2)}$ are constants and $d$ is the depth of the lower layer. The  set of Euler equations is given by
{\small
\begin{align}
\rho_{(i)} (u_{(i)t}+u_{(i)} u_{(i)x}+v_{(i)} u_{(i)y}+w_{(i)}u_{(i)z})+p_{(i)x}=0,\label{2-1}\\
\rho_{(i)} (v_{(i)t}+u_{(i)} v_{(i)x}+v_{(i)}v_{(i)y}+w_{(i)}v_{(i)z})+p_{(i)y}=0,\label{2-2}\\
\varepsilon \rho_{(i)}(w_{(i)t}+u_{(i)} w_{(i)x}+v_{(i)} w_{(i)y}+w_{(i)}w_{(i)z})\nonumber\\
\qquad \qquad +p_{(i)z}+\rho_{(i)}=0,\label{2-3} \\
u_{(i)x}+v_{(i)y}+w_{(i)z}=0,\label{2-4}
\end{align}	
}
where $i=1,2.$ Here, $u$, $v$, $w$ are the velocity components in $x$, $y$, $z$ directions respectively and $p$ is the pressure. 
Using the asymptotic multiple-scales expansions 
\begin{eqnarray*}
\{u_{(i)}, v_{(i)}, w_{(i)} \}=\varepsilon \{\tilde{u}_{(i)}, \tilde{v}_{(i)}, \tilde{w}_{(i)} \}+O(\varepsilon ^2), \\
p_{(i)}=p_{0(i)}(z)+\varepsilon \tilde{p}_{(i)}+O(\varepsilon ^2), \quad \mbox{where }p_{0(i)z}=\rho_{(i)},\\
\rho_{(i)}=\rho_{0(i)}+\varepsilon \tilde{p}_{(i)}+O(\varepsilon ^2),
\end{eqnarray*}
to leading order equations (\ref{2-1}) - (\ref{2-4}) yield
\begin{eqnarray}
\rho_{0(i)} u_{(i)t}+p_{(i)x}=0,\label{2-5}\\
\rho_{0(i)} v_{(i)t}+p_{(i)y}=0,\label{2-6}\\
p_{(i)z}=0,\label{2-7}\\
u_{(i)x}+v_{(i)y}+w_{(i)z}=0.\label{2-8}
\end{eqnarray} 
We apply the free surface and rigid bottom boundary conditions and let $\varepsilon \eta(x,y,t)$ and $\varepsilon \zeta(x,y,t)$ represent the surface and interfacial perturbations, respectively. The boundary and continuity conditions are
\begin{align*}
w_{(1)}=\eta_t, ~~p_{(1)}=\rho_{0(1)}\eta, \quad \mbox{at} ~ z=1+\varepsilon \eta,\\
w_{(2)}=0, \quad \mbox{at} ~ z=0,\\
w_{(1)}=w_{(2)}=\zeta_t,~~p_{(1)}=\zeta\rho_{0(1)},~~p_{(2)}=\zeta\rho_{0(2)}, \\
\quad \mbox{at} ~z=d+\varepsilon \zeta.
\end{align*}
To leading order,
\begin{eqnarray}
&w_{(1)}=\eta_t &\quad \mbox{at}~z=1,\label{2-9}\\
&p_{(1)}=\rho_{0(1)}\eta&\quad \mbox{at}~z=1,\label{2-10}\\
&w_{(2)}=0&\quad \mbox{at}~z=0,\label{2-11}\\
&w_{(1)}=w_{(2)}=\zeta_t&\quad \mbox{at}~z=d,\label{2-12}
\end{eqnarray}
\begin{eqnarray}
&p_{(1)}-p_{(2)}=\zeta(\rho_{0(1)}-\rho_{0(2)})&\quad \mbox{at}~z=d.\label{2-13}
\end{eqnarray}
From equations (\ref{2-5})-(\ref{2-8}) and boundary conditions (\ref{2-9})-(\ref{2-11}), one can obtain
\begin{align*}
w_{(1)t}=(z-1)(\eta_{xx}+\eta_{yy})+\eta_{tt},\\
w_{(2)t}=\frac{z}{\rho_{0(2)}}(p_{(2)xx}+p_{(2)yy}),\\
\quad \mbox{where} ~p_{(2)}=p_{(2)}(x,y,t).
\end{align*}
From the continuity condition (\ref{2-13}) at the interface $z=d$, one gets 
\begin{equation}
p_{(2)}=\rho_{0(1)}\eta+(\rho_{0(2)}-\rho_{0(1)})\zeta.\label{2-14}
\end{equation}
Substituting (\ref{2-14}) into (\ref{2-12}), we obtain
{\small
\begin{align}
&\eta_{tt}= \left (\frac{d(\rho_{0(1)}-\rho_{0(2)})}{\rho_{0(2)}}+1 \right )(\eta_{xx}+\eta_{yy})+\frac{d(\rho_{0(2)}-\rho_{0(1)})}{\rho_{0(2)}}(\zeta_{xx}+\zeta_{yy}),\label{2-15}\\
&\zeta_{tt}=\frac{\rho_{0(1)} d}{\rho_{0(2)}}(\eta_{xx}+\eta_{yy})+\frac{(\rho_{0(2)}-\rho_{0(1)})d}{\rho_{0(2)}}(\zeta_{xx}+\zeta_{yy}).\label{2-16}
\end{align}
}
Now, let us consider the linear combination of $\eta$ and $\zeta$:
$\psi=\eta+b \zeta, \ \mbox{where $b$ is a constant,}$
that satisfies the linear wave equation 
$\psi_{tt}-s^2(\psi_{xx}+\psi_{yy})=0.$
Then, substituting equations (\ref{2-15}) and (\ref{2-16}) into the above equation, we obtain
\begin{align*}
&\quad\psi_{tt}-s^2(\psi_{xx}+\psi_{yy})\\
=& \eta_{tt}+b\zeta_{tt}-s^2(\eta_{xx}+\eta_{yy})-b s^2(\zeta_{xx}+\zeta_{yy})\\
=&\bigg( \frac{d(\rho_{0(1)}-\rho_{0(2)})}{\rho_{0(2)}}+1+\frac{b\rho_{0(1)} d}{\rho_2}-s^2 \bigg) (\eta_{xx}+\eta_{yy})\\
 & \quad+\bigg( \frac{d(\rho_{0(2)}-\rho_{0(1)})}{\rho_{0(2)}}+\frac{b(\rho_{0(2)}-\rho_{0(1)})d}{\rho_{0(2)}}-bs^2\bigg)(\zeta_{xx}+\zeta_{yy})\\
=&0.
\end{align*}
This yields a system of equations 
$$ \left\{ 
 \begin{array}{ll}
   \frac{d(\rho_{0(1)}-\rho_{0(2)})}{\rho_{0(2)}}+1+\frac{b\rho_{0(1)} d}{\rho_{0(2)}}-s^2&=0  ,\\
  \frac{d(\rho_{0(2)}-\rho_{0(1)})}{\rho_{0(2)}}+\frac{b(\rho_{0(2)}-\rho_{0(1)})d}{\rho_{0(2)}}-bs^2&=0   ,
   \end{array} \right.$$
which has two solutions
$$ \left\{ 
 \begin{array}{ll}
   s_{+}^2 &=
   \frac{1+\sqrt{D}}{2},\\
   b_1 &= \left (-2d(\rho_{0(1)}-\rho_{0(2)})-\rho_{0(2)}(1-\sqrt{D}) \right ) / 2d\rho_{0(1)} ,
   \end{array} \right.$$
$$ \left\{ 
 \begin{array}{ll}
   s_{-}^2 &=
   \frac{1-\sqrt{D}}{2},\\
   b_2 &=\left (2d(\rho_{0(1)}-\rho_{0(2)})-\rho_{0(2)}(1+\sqrt{D}) \right ) / 2d\rho_{0(1)} ,
   \end{array} \right.$$
where $D=(1-2d)^2+\frac{4d(1-d)\rho_{0(1)}}{\rho_{0(2)}}.$

The wave speeds $s_+$ and $s_-$ coincide with the speeds of the surface and interfacial modes in the absence of a shear flow in Section II.
The modal equations  in $(x,y,z)$ coordinates in the two-layer case have the form
\begin{eqnarray}
s^2\phi_{zz}&=&0\quad \mbox{at}~0<z<1,\label{mm1}\\
s^2\phi_z-\phi&=&0\quad \mbox{at}~z=1,\label{mm2}\\
\phi&=&0\quad \mbox{at}~z=0.\label{mm3}
\end{eqnarray} 
We obtain the modal function $\phi$ in the form
\begin{align*}
 \phi=\left\{ 
 \begin{array}{ll}
  \Lambda(s^2-1+z)\quad \mbox{at}~d<z<1,\\
  (s^2-1+d)\frac{\Lambda z}{d} \quad \mbox{at}~0<z<d,
   \end{array} \right.
   \end{align*}
where $\Lambda$ is a constant. For the surface mode, the wave speed $s=s_{+}$ and we set $\phi_s=1$ at $z=1$, which implies $\Lambda=1/s_{+}^2$. The modal function for the surface mode is
$$ \phi_s=\left\{ 
 \begin{array}{ll}
  \frac{s^2_{+}-1+z}{s_{+}^2}\quad \mbox{at}~d<z<1,\\
  \frac{(s^2_{+}-1+d)z}{s_{+}^2 d} \quad \mbox{at}~0<z<d.
   \end{array} \right.$$
For the interfacial mode, the wave speed $s=s_{-}$ and we set $\phi_s=1$ at $z=d$, which implies $\Lambda=1/(s^2_{-}-1+d)$. The modal function for the interfacial mode is
$$ \phi_i=\left\{ 
 \begin{array}{ll}
  \frac{s^2_{-}-1+z}{s^2_{-}-1+d}\quad \mbox{at}~d<z<1,\\
  \frac{z}{ d} \quad \mbox{at}~0<z<d.
   \end{array} \right.$$
Then,
\begin{eqnarray*}
\zeta=\left( A_s\phi_s+A_i\phi_i\right)_{z=d},\qquad \eta=\left(A_s\phi_s+A_i\phi_i\right)_{z=1},
\end{eqnarray*}
where $A_s$ denotes the wave amplitude of the surface waves at $z=1$ and $A_i$ denotes the wave amplitude of the interfacial waves at $z=d$.
Thus,
\begin{align*}
\psi_1=\eta+b_1\zeta \qquad\qquad \qquad \qquad\qquad\qquad\qquad\\
=A_s+A_i\frac{s_{-}^2}{s_{-}^2-1+d}+b_1 \frac{s_{+}^2-1+d}{s_{+}^2}A_s+b_1A_i\\
=P_{1s}A_s+P_{1i}A_i,\qquad\qquad\qquad\qquad\qquad\qquad
\end{align*}
where
\begin{align*}
P_{1s}=1+\frac{\rho_2}{d\rho_1} \cdot \frac{\rho_2s^2_{+}-(1-d)\rho_2-d\rho_1}{s_{+}^4}\cdot (s_{+}^2-1+d)~~~\\
  =   1+\frac{\rho_2(\sqrt{D}+(2d-1))(\rho_2\sqrt{D}+(2d-1)\rho_2-2d\rho_1)}{d\rho_1(1+\sqrt{D})^2}  ,
  \end{align*}
  $$P_{1i}=\frac{s_{-}^2}{s_{-}^2-1+d}+\frac{\rho_2}{d\rho_1}\big(s_{+}^2-\frac{(1-d)\rho_2+d\rho_1}{\rho_2}\big)=0,$$
and
\begin{align*}
(A_s)_{tt}-s_+^2((A_s)_{xx}+(A_s)_{yy})=\frac{\psi_{1tt}-s_+^2(\psi_{1xx}+\psi_{1yy})}{P_{1s}}=0.
\end{align*}
Similarly, one can show that $\psi_2=P_{2i}A_i$, where $P_{2i}$ is a constant, and the interfacial mode $A_i$ also satisfies the linear wave equation
$$(A_i)_{tt}=s_-^2((A_i)_{xx}+(A_i)_{yy}).$$


\section{Numerical method}
The cKdV-type equation (\ref{cKdV}) is written in the form
\begin{equation} 
\mu_1 A_R+\mu_2 AA_\xi+\mu_3 A_{\xi\xi\xi}+\mu_4 \frac{A}{R}+\mu_5 \frac{A_\theta}{R}=0, \label{A1}
\end{equation}
where $\mu_i=\mu_i(\theta),~i=\overline{1..5}.$
A finite-difference scheme used in this paper is an extension of the scheme suggested by Feng and Mitsui \cite{Feng98}.

\subsection{Linearized implicit method}
We assume that $\xi \in [\xi_{\text{min}},\xi_{\text{max}}], R \in [R_0, R_{\text{max}}]$ and $\theta \in [0, 2 \pi]$.
We discretise the domains of $\xi, R$ and $\theta$ into grids with equal spacings $\Delta \xi, \Delta R$ and $\Delta\theta$. 
We approximate the grid values 
$A(\xi_{\text{min}}+l \Delta \xi,~R_0 + n \Delta R,~ m \Delta \theta)$ by $A_{l,m}^n$, where 
$l=0,1,2,...,L;~m=0,1,2,...,M;~n=0,1,2,...,N$ with $L=(\xi_{\text{max}}-\xi_{\text{min}})/\Delta \xi$ and $M=2\pi/\Delta \theta-1$ ($\theta=0$ and $\theta=2\pi$ define the same direction), and approximate the 
coefficients   $\mu_i(m \Delta \theta)$ by $\mu_{i,m}$. The initial condition is given by  $\bm{A^0}=[A_{11}^0,A_{21}^0,\dots, A_{L1}^0,A_{12}^0,\dots, A_{LM}^0]^T$. 

The central difference approximations of partial derivatives in (\ref{A1}) are 
\begin{align}
\left.A_\xi\right|_{l,m}^n = \frac{A_{l+1,m}^n-A_{l-1,m}^n}{2\Delta \xi}+O(\Delta\xi^2),\qquad\qquad\qquad\qquad \label{A2}\\
\left.A_{\xi\xi\xi}\right|_{l,m}^n = \frac{A_{l+2,m}^n-2A_{l+1,m}^n+2A_{l-1,m}^n-A_{l-2,m}^n}{2\Delta \xi^3 } \qquad \qquad   \nonumber \\
+O(\Delta \xi^2), \qquad \qquad \qquad \label{A3} \\
\left. A_\theta \right|_{l,m}^n = \frac{A_{l,m+1}^n-A_{l,m-1}^n}{2\Delta \theta}+ O(\Delta \theta^2).\qquad\qquad\qquad\qquad \label{A4}
\end{align}
We denote $f=A^2$ and $\frac 12 f_\xi=AA_\xi$. A set of nonlinear algebraic equations has to be solved in order to obtain $ \bm{A^{n+1}}$ from $\bm{A^n}$. We need to linearise the equations using the Taylor expansion for $f$:
\begin{eqnarray*}
f_{l,m}^{n+1}= f_{l,m}^n+\left. \frac{\partial f}{\partial R} \right|_{l,m}^n \Delta R +O(\Delta R^2) \\
= f_{l,m}^n+ D_{l,m}^n \Delta A_{l,m}^{n+1}+O(\Delta R^2),
\end{eqnarray*}
where $D_{l,m}^n =\left. \frac{\partial f}{\partial A} \right|_{l,m}^n =2 A_{l,m}^n$ and $\Delta A_{l,m}^{n+1}=A_{l,m}^{n+1}-A_{l,m}^n$. Then 
$$f_{l,m}^{n+1}+f_{l,m}^n\approx 2f_{l,m}^n+2 A_{l,m}^n(A_{l,m}^{n+1}-A_{l,m}^n)=2A_{l,m}^nA_{l,m}^{n+1}.$$

Using the central difference approximations, the equation (\ref{A1}) at $R_n+\frac 12 \Delta R$ can be written as 
{\small
\begin{eqnarray}
\mu_{1,m} \frac{A_{l,m}^{n+1}-A_{l,m}^n}{\Delta R}+ \mu_{2,m}\frac{(A_{l,m}^n A_{l,m}^{n+1})_\xi}2 +\mu_{3,m} \frac{(A_{l,m}^{n+1}+A_{l,m}^n)_{\xi\xi\xi}} 2\nonumber \\
+\frac{\mu_{4,m}}{2}\left (\frac{A_{l,m}^n}{R_n}+\frac{A_{l,m}^{n+1}}{R_{n+1}}\right )+ \frac{\mu_{5,m}}{2} \left (\frac{(A_{l,m}^n)_\theta}{R_n}+\frac{(A_{l,m}^{n+1})_\theta}{R_{n+1}}\right )=0.\label{4-4}
\end{eqnarray}
}
Substituting (\ref{A2})-(\ref{A4}) into the above equation we obtain the following linear system of equations:
{\small
\begin{eqnarray}
\frac{\mu_{5,m}}{4R_{n+1}\Delta \theta}A_{l,m+1}^{n+1}-\frac{\mu_{5,m}}{4R_{n+1}\Delta \theta} A_{l,m-1}^{n+1}+\frac{\mu_{3,m}}{4\Delta\xi^3}A_{l+2,m}^{n+1}\nonumber\\
+\big(\frac{\mu_{2,m}}{4\Delta \xi}A_{l+1,m}^n-\frac{\mu_{3,m}}{2\Delta \xi^3}\big)A_{l+1,m}^{n+1}+\big(\frac{\mu_{1,m}}{\Delta R}+\frac{\mu_{4,m}}{2R_{n+1}}\big)A_{l,m}^{n+1}\nonumber \\
+\big(-\frac{\mu_{2,m}}{4\Delta \xi}A_{n-1,m}^n+\frac{\mu_{3,m}}{2\Delta\xi^3}\big)A_{l-1,m}^{n+1}-\frac{\mu_{3,m}}{4\Delta \xi^3}A_{l-2,m}^{n+1}=d_{l,m}^n~,\label{A5}
\end{eqnarray}
}
where
\begin{align*}
d_{l,m}^n= -\frac{\mu_{5,m}}{4R_n\Delta\theta}A_{l,m+1}^n+\frac{\mu_{5,m}}{4R_n \Delta \theta}A_{l,m-1}^n-\frac{\mu_{3,m}}{4\Delta\xi^3}A_{l+2,m}^n\\
+\frac{\mu_{3,m}}{2\Delta\xi^3}A_{l+1,m}^n+\big(\frac{\mu_{1,m}}{\Delta R}-\frac{\mu_{4,m}}{2R_n}\big)A_{l,m}^n-\frac{\mu_{3,m}}{2\Delta\xi^3}A_{l-1,m}^n\\
+\frac{\mu_{3,m}}{4\Delta\xi^3}A_{l-2,m}^n.
\end{align*}
Equation (\ref{A5}) can be written in the vector form: 
\begin{equation}
\hspace{1cm} \bm{T}\cdot\bm{A^{n+1}}=\bm{d^{n}}.
\label{vfe}
\end{equation}

At the boundary, using periodicity of $\theta$, we have:
\begin{eqnarray}
A_{l,-1}^n=A(\xi,R,-\Delta\theta)=A(\xi,R,2\pi-\Delta\theta)=A_{l,M}^n~,   \nonumber\\
A_{l,M+1}^n=A(\xi,R,2\pi)=A(\xi,R,0)=A_{l,0}^n~,
\label{abc1}
\end{eqnarray}
for every $n$ and $l$. 

The domain of $\xi$ is chosen to satisfy the condition that $A_{l,m}^n$ tends to $0$ at $l=0$ and $L$, i.e. at $\xi=\xi_{\text{min}}$ and  $\xi_{\text{max}}$. Then the value of $A_{l,m}^n$ outside of the interval $[\xi_{\text{min}},\xi_{\text{max}}]$ is equal to  $0$,
\begin{equation}
A_{-2,m}^n=A_{-1,m}^n=A_{L+1,m}^n=A_{L+2,m}^n=0.
\label{abc2}
\end{equation}
From equation (\ref{A5}), the terms in the matrix of coefficients $\bm{T}$ are determined by the following formulae
\begin{eqnarray*}
&&a_{lm,l(m+1)}^n =\frac{\mu_{5,m}}{4R_{n+1}\Delta\theta},\\
&&a_{lm,l(m-1)}^n =-\frac{\mu_{5,m}}{4R_{n+1}\Delta\theta},\\
&&a_{lm,(l+2)m}^n =\frac{\mu_{3,m}}{4\Delta \xi^3},\\
&&a_{lm,(l+1)m}^n = \frac{\mu_{2,m}}{4\Delta \xi}A_{l+1,m}^n-\frac{\mu_{3,m}}{2\Delta\xi^3},\\
&&a_{lm,lm}^n = \frac{\mu_{1,m}}{\Delta R}+\frac{\mu_{4,m}}{2R_{n+1}},
\end{eqnarray*}
\begin{eqnarray*}
&&a_{lm,(l-1)m}^n = -\frac{\mu_{2,m}}{4\Delta \xi}A_{l+1,m}^n+\frac{\mu_{3,m}}{2\Delta\xi^3},\\
&&a_{lm,(l-2)m}^n = -\frac{\mu_{3,m}}{4\Delta \xi^3}.
\end{eqnarray*}
Note that the matrix of coefficients at the boundary needs to be changed in accordance with the boundary conditions (\ref{abc1}) and (\ref{abc2}). 
The coefficients $\mu_i$ are determined by the formulae (\ref{c2}-\ref{c5}). 

\subsection{Order of accuracy}

We use the  Taylor expansions of $A_{l,m}^{n+1}$ and $A_{l,m}^{n+1}/R^{n+1}$:
\begin{align*}
A_{l,m}^{n+1}=A_{l,m}^n+\Delta R (A_{l,m}^n)_R+\frac 12 (\Delta R)^2 (A_{l,m}^n)_{RR}+O(\Delta R^3),\\
\frac{A_{l,m}^{n+1}}{R^{n+1}}=\frac{A_{l,m}^n}{R^n}+\Delta R \left (\frac{A_{l,m}^n}{R^n}\right )_R+O(\Delta R^2).\qquad\quad\qquad\qquad 
\end{align*}
 Substituting the central difference approximations (\ref{A2})-(\ref{A4}) and the Taylor expansions above into the difference equation (\ref{4-4}), we obtain
{\small \begin{align}
\mu_{1,m}((A_{l,m}^n)_R+\frac 12 \Delta R (A_{l,m}^n)_{RR})+\frac{\mu_{2,m}}{2}\left (2f_{l,m}^n+\Delta \left (\frac{\partial f}{\partial R}\right )_{l,m}^n\right )_\xi \nonumber\\
+\frac{\mu_{3,m}}{2}(2A_{l,m}^n+\Delta R(A_{l,m}^n)_R)_{\xi\xi\xi}+\frac{\mu_{4,m}}{2}\left (\frac{2A_{l,m}^n}{R^n}+\Delta R \left (\frac{A_{l,m}^n}{R_n}\right )_R\right )\nonumber\\
+\frac{\mu_{5,m}}{2} \left (2\left (\frac{A_{l,m}^n}{R_n}\right )_\theta+\Delta R\left (\frac{A_{l,m}^n}{R_n} \right )_{R\theta}\right )+O(\Delta R^2+\Delta \xi^2,\Delta R^2 +\Delta \theta^2)\nonumber\\
=\mu_{1,m}(A_{l,m}^n)_R+\mu_{2,m}(f_{l,m}^n)_\xi+\mu_{3,m}(A_{l,m}^n)_{\xi\xi\xi}+\mu_{4,m}\frac{A_{l,m}^n}{R_n}\nonumber\\
+\mu_{5,m}\left (\frac{A_{l,m}^n}{R_n}\right )_\theta  +\frac 12 \Delta R \left (  \mu_{1,m} (A_{l,m}^n)_R+\mu_{2,m}(f_{l,m}^n)_\xi+\mu_{3,m}(A_{l,m}^n)_{\xi\xi\xi} \right .\nonumber\\
\left .  +\mu_{4,m}\frac{A_{l,m}^n}{R_n}+\mu_{5,m}\left (\frac{A_{l,m}^n}{R_n}\right )_\theta  \right)_R +O(\Delta R^2+\Delta \xi^2,\Delta R^2 +\Delta \theta^2).\label{A6}
\end{align}}
If $A_{l,m}^n$ is an exact solution of the cKdV - type equation (\ref{A1}), then
{\small
$$\mu_{1,m}(A_{l,m}^n)_R+\mu_{2,m}(f_{l,m}^n)_\xi+\mu_{3,m}(A_{l,m}^n)_{\xi\xi\xi}+\mu_{4,m}\frac{A_{l,m}^n}{R_n}+\mu_{5,m}\left (\frac{A_{l,m}^n}{R_n}\right )_\theta=0,$$
}
and the truncation error of 
the system (\ref{A5}) is $O(\Delta R^2+\Delta \xi^2,\Delta R^2 +\Delta \theta^2)$.

\subsection{Physical coordinates}
The wave amplitude $A$  in equation (\ref{A1}) depends on the variables $(\xi,R,\theta)$. In physical coordinates, it depends on the radius $r$, the time $t$ and the angle $\theta$. The two coordinate systems are related as follows:
$$
\xi=rk(\theta) -st,\quad  R=\varepsilon r k(\theta),\quad  \theta=\theta,
$$
$$
\Longrightarrow
r=\frac{R}{\varepsilon k(\theta)},\quad  t=\frac{R}{\varepsilon s}-\frac{\xi}{s},\quad  \theta=\theta,
$$
where $\varepsilon$ is the amplitude parameter, $s$ is the wave speed in the absence of a shear flow and the function $k(\theta)$ is the `distortion function' ($k(\theta) = 1$ in the absence of a shear flow).
The range and discretisation of the variables $\xi, R$ and $\theta$ is discussed in Appendix C.1.

The initial condition for the derived equation (\ref{A1}) is given at fixed $R_0$ in the form $A(R_0, \xi, \theta)$ . In an experiment, one can take probes at fixed points to measure the wave amplitude at various depths.
This method has been used, for example, by Ramirez et al \cite{ramirez2002}. The initial condition  (at $R = R_0= const$) can be obtained by taking probes at points $r_0=R_0/ \varepsilon k(\theta)$
and measuring the wave amplitude in the required time interval $t\in[0,t_1]$. Here
${\displaystyle t_1=\frac{R_0}{\varepsilon s}-\frac{\xi_{\text{min}}}{s},}$
which implies
${\displaystyle \xi_{\text{min}}=\frac{R_0}{\varepsilon}-st_1.}$ Similarly, $\xi_{\text {max}} = \frac{R_0}{\varepsilon}$. Thus, ${\displaystyle \xi\in [R_0/\varepsilon-st_1,R_0/\varepsilon].}$
We assume that waves are generated by a disturbance in some neighbourhood of the origin ($r < r_0$). Therefore,
we impose the conditions  $A_{l,m}^n=0$  for 
$l>((R_0 + n\Delta R) /\varepsilon-\xi_{\text{min}})/\Delta\xi$. 


For a given point $(r,\theta)$, the values of $m$ and $n$ used to calculate the quantities $A_{l,m}^n$ are given by
${\displaystyle m=\frac{\theta}{\Delta \theta},\ n=\frac{\varepsilon k(\theta)(r-r_0)}{\Delta R},}$
and we change the value of $l$ as a function of $t$:
${\displaystyle l=\frac{rk(\theta)-\xi_{\text{min}}-st}{\Delta \xi}.}$
This map allows us plot the wave profile at a given moment of time $t$ in the coordinates $(r,\theta)$. Note that the model is used to find the wave amplitude in the area $r>r_0$.  We showed the solution of the linear problem (defining our initial condition at $r=r_0$)  in the area $r<r_0$.




\begin{thebibliography}{99}

\bibitem[Maxon $\&$ Viecelli (1974)]{Maxon74} S. Maxon,  J. Viecelli, Cylindrical solitons,  Phys. Fluids 17 (1974) 1614 - 1616.

\bibitem[Miles (1978)]{Miles78} J.W. Miles, An axisymmetric Boussinesq wave, J. Fluid Mech. 84 (1978) 181 - 191.

\bibitem[Johnson (1980)]{Johnson80} R.S. Johnson, Water waves and Korteweg - de Vries equations, J. Fluid Mech. 97 (1980) 701 - 719.

 \bibitem[Dorfman et al. (1981)]{Dorfman81} A.A. Dorfman, E.N. Pelinovskii, Yu.A. Stepanyants, Finite-amplitude cylindrical and spherical waves in weakly dispersive media, 
Sov. Phys. J. Appl. Mech. Tech. Phys. 2 (1981) 206 - 211.

\bibitem[Stepanyants (1981)]{Stepanyants81} Yu.A. Stepanyants, Experimental investigation of cylindrically diverging solitons in an electric lattice, Wave Motion 3 (1981) 335 - 341. 

\bibitem[Weidman $\&$ Zakhem (1988)]{Weidman88} P.D. Weidman, R. Zakhem, Cylindrical solitary waves, J. Fluid Mech. 191 (1988) 557-573.



\bibitem[Lipovskii (1985)]{Lipovskii85} V.D. Lipovskii, On the nonlinear internal wave theory in fluid of finite depth, Izv. Akad. Nauk SSSR, Ser. Fiz. Atm. Okeana 21 (1985) 864 - 871.

\bibitem[Johnson (1990)]{Johnson90} R.S. Johnson, Ring waves on the surface of shear flows: a linear and nonlinear theory, J. Fluid Mech. 215 (1990) 145 - 160.

\bibitem[Johnson (1997)]{Johnson_book} R.S. Johnson, A modern introduction to the mathematical theory of water waves, 
Cambridge University Press, Cambridge, 1997.


\bibitem[Khusnutdinova $\&$ Zhang(2014)]{kx14} K.R. Khusnutdinova, X. Zhang, Long ring waves in a stratified fluid over a shear flow, J. Fluid Mech. 794 (2016) 17-44.

 \bibitem[Druma (1976)]{Druma76} V.S. Druma, Analytical solution of the axially symmetric KdV equation, Izv. Akad. Nauk MssR 3 (1976) 14 - 16 (in Russian).

\bibitem[Calogero $\&$ Degasperis (1978)]{Calogero78} F. Calogero, A. Degasperis, Solution by the spectral transform method of a nonlinear evolution equation including as a special case the cKdV equation, Lett. Nuovo Cim. 23 (1978) 150 - 154. 



\bibitem[Burns (1953)]{Burns53} J.C. Burns, Long waves in running water, Proc. Camb. Phil. Soc. 49 (1953) 695 - 706.


\bibitem[Feng and Mitsui (1998)]{Feng98}
B.F. Feng, T. Mitsui,  A finite difference method for the Korteweg - de Vries and the Kadomtsev - Petviashvili equations, J. Comp. Appl. Math.  90 (1998) 95 - 116.

\bibitem[Dobrokhotov (2010)]{Dobrokhotov}
S.Y. Dobrokhotov, S.Y. Sekerzh-Zen'kovich, A class of exact algebraic localised solutions of the multidimensional wave equation, Math. Notes 88 (2010) 894 - 897.

\bibitem[Gurevich and Pitaevskii (1974)]{gurevich1974} A.V. Gurevich, L.P. Pitaevskii, Nonstationary structure of a collisionless shock wave, Sov. Phys. JETP 38 (1974) 291 - 297.


\bibitem[Hoefer and Ablowitz (2009)] {hoefer2009} M. Hoefer, M. Ablowitz, Dispersive shock waves,  Scholarpedia, 2009. 

\bibitem[Hoefer et. al. (2006)] {hoefer2006} M.A. Hoefer, M.J. Ablowitz, M.J., I. Coddington, E.A. Cornell, P. Engels, V. Schweikhard, Dispersive and classical shock waves in Bose-Einstein condensates and gas dynamics, Phys. Rev. A. 74 (2006) 023623.

\bibitem[Kamchatnov et. al. (2004)] {kamchatnov2004} A.M. Kamchatnov, A. Gammal, R.A. Kraenkel, Dissipatiionless shock waves in Bose-Einstein condensates with repulsive interaction between atoms, Phys. Rev. A. 69 (2004) 063605.

\bibitem[Whitham (1999)] {whitham_book} G.B. Whitham, Linear and nonlinear waves, Wiley, New York, 1999. 

\bibitem[Smyth and Holloway (1988)] {smyth1988} N.F. Smyth, P.E. Holloway, Hydraulic jump and undular bore formation on a shelf break, J. Phys. Oceanogr. (1988) 947 - 962.

\bibitem[El et. al. (2007)] {el2007} G.A. El, R.H.J. Grimshaw, A.M. Kamchatnov, Evolution of solitary waves and undular bores in shallow-water flows over a gradual slope with bottom friction, J. Fluid. Mech. 585 (2007) 213 - 244.

\bibitem[Ablowitz et. al. (2015)] {ablowitz2015} M.J. Ablowitz, Ali Demirci, Ali,  Yi-Ping Ma, Dispersive shock waves in the Kadomtsev-Petviashvili and Two Dimensional Benjamin-Ono equations, arXiv:1507.08207v1 (2015).

\bibitem[Esler and Pearce (2011)] {esler2011} J.G. Esler, J.D. Pearce, Dispersive dam-break and lock exchange flows in a two-layer fluid, J. Fluid Mech. 667 (2011) 555 - 585.

\bibitem[Chumakova et. al. (2009)] {Chumakova} L. Chumakova, F.E. Menzaque, P.A. Milewski, R.R. Rosales, E.G. Tabak, C.V. Turner, Stability properties and nonlinear mappings of two and three-layer stratified flows, Stud. Appl. Math. 122 (2009)123 - 137.

\bibitem[Grue \& Sveen (2010)] {grue} J. Grue, J.K. Sveen, A scaling law of internal run-up duration, Ocean Dynamics 60 (2010) 993-1006. 

\bibitem[Arkhipov et. al. (2013)]{ark2013} D.G. Arkhipov, G.A. Khabakhpashev,  N.S. Safarova, Simulation of moderately long nonlinear spatial waves on the interface between two fluid flows in a horizontal channel, Eur. J. Mech. - B/Fluids 39 (2013) 87 - 94.

\bibitem[Arkhipov et. al. (2015) ]{arkhipov2015} D.G. Arkhipov, GA. Khabakhpashev, V.E. Zakharov, Describing dynamics of nonlinear axisymmetric waves in dispersive media with new equation, Phys. Lett.  379 (2015) 1414 - 1417.

\bibitem[Ellingsen (2014)]{ellingsen14} S.A. Ellingsen, Initial surface disturbance on a shear current: The Cauchy-Poisson problem with a twist, Phys. Fluids 26 (2014) 082104. 

\bibitem[Li \& Ellingsen (2016)] {li} Y. Li, S.A. Ellinsen, Water waves from general, time-dependent surface pressure distribution in the presence of a shear current,  to appear in Int. J. Offshore Polar Eng. (2016)

\bibitem[LeBlond and Mysak (1978)] {leblond} P.H. LeBlond, L.A.  Mysak, Waves in the ocean, Elsevier, Amsterdam, 1978.

\bibitem[C. Ramirez, D. Renouard, and Yu.A. Stepanyants  (2002)]{ramirez2002}
C. Ramirez, D. Renouard, Yu. A. Stepanyants, Propagation of cylindrical waves in a rotating fluid, Fluid Dyn. Res. 30 (2002) 169 - 196.


\end{thebibliography}
\end{document}